\documentclass[final,5p,times]{elsarticle}
\usepackage{lineno}
\usepackage{soul}
\usepackage{graphicx}
\usepackage{amssymb}
\usepackage{verbatim}
\usepackage{multirow}
\usepackage{tabularx, colortbl}
\usepackage[hidelinks]{hyperref}
\hypersetup{
    colorlinks = true,
    linkcolor = blue,
    anchorcolor = blue,
    citecolor = blue,
    filecolor = blue,
    urlcolor = blue
    }
\usepackage{gensymb}
\usepackage{csquotes,textcomp}
\usepackage{amssymb,graphicx,amsmath, amsthm,verbatim}

\usepackage{color}
\definecolor{amethyst}{rgb}{0.6, 0.4, 0.8}
\definecolor{alizarin}{rgb}{0.82, 0.1, 0.26}
\definecolor{green}{rgb}{0.55, 0.71, 0.0}
\definecolor{apricot}{rgb}{0.98, 0.81, 0.69}
\definecolor{auburn}{rgb}{0.43, 0.21, 0.1}
\definecolor{babyblueeyes}{rgb}{0.63, 0.79, 0.95}
\definecolor{bittersweet}{rgb}{1.0, 0.44, 0.37}
\definecolor{arsenic}{rgb}{0.23, 0.27, 0.29}
\usepackage[normalem]{ulem} % for strikethrough
\newcommand{\cfsout}{\bgroup\markoverwith{\textcolor{red}{\rule[0.5ex]{2pt}{0.4pt}}}\ULon}

\begin{document}
\begin{frontmatter}

   \title{A study on Performance Boost of a 17~m class  \\ Cherenkov telescope with a SiPM-based camera} 

  \author{Cornelia~Arcaro$^{1~\dagger}$, Michele~Doro$^{1,2~\dagger}$, Julian Sitarek$^{3~\dagger}$, Dominik Baack$^4$, }
  
  \address{
    $^1$ Istituto Nazionale di Fisica Nucleare (INFN) sez. Padova, via Marzolo 8, I-35131,
    Padova (Italy)\\
    $^2$
    University of Padova, Department of Physics and Astronomy, via Marzolo 8, I-35131, Padova (Italy)\\
    $^3$ University of Lodz, Faculty of Physics and Applied Informatics, Department of Astrophysics, 90-236 Lodz, Poland\\
    $^4$ Technische Universit\"at Dortmund, D-44221 Dortmund, Germany\\
    $\dagger$ Correspondence: cornelia.arcaro@pd.infn.it, michele.doro@unipd.it,
    jsitarek@uni.lodz.pl 
  }

  \begin{abstract}
The current generation of Imaging Atmospheric Cherenkov Telescopes (IACTs), comprised of major installations such as the MAGIC, H.E.S.S.  and VERITAS telescopes, is sometimes called the 3$^{\mathrm{rd}}$ generation of such instruments. These telescopes use multipixel cameras composed of hundreds up to thousands photomultiplier tubes (PMTs). 
%\mdc{The next sentence is in fact partial and maybe useless, it can be removed:} The total light yield recovered by such instruments depends, besides the PMT photon detection efficiency (PDE), on the mirror dish reflectivity, the light absorption by the camera window, as well as the electronics. 
The supremacy of PMTs is currently being challenged by photon sensors, rapidly spreading in popularity: the silicon photomultipliers (SiPMs), that are becoming a valid alternative thanks to their high PDE, low operating voltage and flexibility in installation. In this report, we investigate the performance of an existing 3$^{\mathrm{rd}}$ generation IACT array (taking as a working example MAGIC), in which PMTs would be replaced with SiPMs, applying generalized simulations, not tuned for a specific hardware solution. 
%We find an increase of sensitivity up to a factor of 2 for energies below 200~GeV, which corresponds to a factor of 4 less observation time. 
%We find an energy-dependent improvement in sensitivity, reaching a factor of two-three at the current trigger threshold energy. 
We find an energy-dependent improvement in sensitivity, reaching a factor of three at the current trigger threshold energy. 
Interestingly, we also find that the stronger sensitivity of SiPMs in the red part of the spectrum, a source of background for IACTs, does not affect this performance.
\end{abstract}


\end{frontmatter}

\section{Introduction}
\label{sec:intro}

Imaging Atmospheric Cherenkov Telescopes (IACTs) detect the
  Cherenkov light radiated in Extensive
  Atmospheric Showers (EASs) in the Earth's atmosphere at the height of 10-20~km a.s.l \citep{Hillas:2013txa,Lorenz:2012nw}. EASs are produced by energetic cosmic rays or gamma
  rays impinging the Earth. The Cherenkov radiation is peaked in the blue-UV, has a duration of few nanoseconds, and illuminates an area of few tens of thousands square meters at ground level, depending on the energy of the primary particle. The observations can only be performed at moonless nights or nights with moderate moonlight and are made through telescopes with large light collectors. However, in recent times, technical solutions were implemented to operate also with stronger and stronger moon levels, selecting targets at a sufficient angular distance from the Moon \citep[see, e.g.,][]{MAGIC:2017zph,Griffin:2015zsa}. For concreteness, we base our study on a specific 3$^{\mathrm{rd}}$ generation IACT system, the Major Atmospheric Gamma-ray Imaging Cherenkov (MAGIC) telescopes, operating since 2003 at the Observatorio Roque de Los Muchachos (ORM) on the island of La Palma. In \autoref{sec:discussion}, we discuss about the applicability of these results for other installations.
  %, consisting of two instruments
  Each of the two MAGIC telescopes has a 17~m-diameter ($f-$number$=1$) parabolic reflector, tessellated with about 250 mirror facets~\cite{Doro:2008zz}.  To catch the brief Cherenkov light flashes, the MAGIC telescopes are equipped with two identical cameras, each composed of 1039 photomultiplier tubes (PMTs).
  The PMTs are organised in $265$ clusters of 7~photodetectors each, assembled in modules that can easily be plugged in or out independently from each other.
  It is therefore possible, to separately remove a module, and replace it with a different one, provided the front- and back-end electronic is compatible.
  The MAGIC PMTs are Hamamatsu R10408 6-dynodes with an hemispherical bi-alkali cathode of $1'$ diameter, operated at a voltage of $\approx 1$~kV. Each PMT is equipped with a compound parabolic concentrator adapted from the design of the so-called Winston cone~\citep{winston_cone,winstoncone}, that reduces the dead area between neighbouring PMTs and avoids the stray-light coming from far-off the mirror dish.  
  Silicon Photomultipliers (SiPMs) are photosensors composed of microscopic diode cells assembled in matrices of thousands, to reach sizes of few mm across, discussed in more details in \autoref{sec:sipm}. They are currently replacing PMTs where compactness is required and because they operate at low voltage. 
  These devices are advancing rapidly in the field of fast photodetectors. Compared to PMTs, besides their low operating voltage, they provide advantages such high efficiencies, insensitivity to magnetic fields, and robustness against bright ambient light. The compact geometrical size of SiPMs allow to easily arrange them, but makes it complicate to cover large areas such as those of IACT cameras, of the order of two square meter. Other drawbacks compared to Cherenkov applications of PMTs are the optical crosstalk between adjacent cells and the larger sensitivity to red light, as well as a higher dark current. Normally multiple SiPMs must be combined together to match the sensitive area of a PMT, in this case, also the signal summation could provide larger time-spread signal than that of PMTs. 
  %Currently, the development and testing of  SiPM-based modules in the MAGIC cameras~\cite{hahn:2022} is ongoing and the possibility of replacing a part or the whole MAGIC cameras with modules based on SiPMs is under consideration. 
  Currently, the development and testing of  SiPM-based modules in the MAGIC cameras is ongoing in preparation of a possible camera upgrade for MAGIC or the future Cherenkov Telescope Array \cite{2018NIMPA.912..259H}. 
Furthermore, SiPMs are currently already adopted as main photosensors in some telescope designs for the Cherenkov Telescope Array (CTA) \cite{Doro:2009qs, Consortium:2010bc,Acharya:2013sxa}. This solution is valid especially in the case of a double-optics Schwarzschild-Couder design in which a secondary mirror not only counteracts some geometrical aberrations, but also allows for a reduction of the camera occupancy and therefore the size of its pixels~\citep{Heller:2016rlc}. There are also studies to equip single-optics Davies-Cotton or parabolic CTA telescopes with SiPMs, but so far only the technological feasibility is under evaluation~\cite{Rando:2015jpa, Arcaro:2017owo, Mallamaci:2018ago, perennes:2020}.

  \medskip
  It is therefore interesting to evaluate the expected performance of the 3$^{\mathrm{rd}}$-generation IACTs with SiPM-based cameras. 
  %IT WAS A REPETITION. This work uses as case study the MAGIC telescopes, however, similar reasonable conclusions may be found for other current-generation instruments, as discussed in \autoref{sec:discussion}. 
  However, a complete estimation of the validity of a SiPM-equipped camera for this generation of telescopes is out of the scope of this paper. This would require a) comparing selected replacement of photosensors and electronics to a complete new assembly of the camera b) evaluating in details the grouping of SiPMs into single pixels, c) using specifically developed SiPMs after bidding with companies, as well as several other aspects, not excluding the related cost and scientific opportunity. In this work, we focus on a `minimal hardware change' scenario, in which we replaced the MAGIC PMT modules with clusters of SiPMs, leaving the rest of the system untouched. This approach allows us to give a valuable input for the expected performance of a new SiPM-based camera.

   \medskip
 The paper is structured as follows: In \autoref{sec:sipm}  we discuss the focal-plane photosensors, the current generation PMTs and SiPMs, comparing their performance. In \autoref{sec:yield3} we review the factors that affect the light yield of the camera, especially considering how the original spectral distribution of the Cherenkov light is modified through the detection by the instrument. In \autoref{sec:yield4} we compute the light yield for the case of PMTs and SiPMs. We also discuss the possibility to filter out the red part of the spectrum with a dedicated dichroic window in front of the camera. 
 The details of the SiPM camera simulations are presented in \autoref{sec:simulation}.
 The performance with a SiPM-based camera is discussed in \autoref{sec:sensitivity}. In \autoref{sec:discussion} we  discuss the technical feasibility of a potential camera upgrade and the accuracy of our results. We summarise and conclude in \autoref{sec:conclusion}.

\section{Focal plane photosensors}
\label{sec:sipm}
SiPMs are grids of Single Photon Avalanche photoDiodes (SPADs) on silicon
substrate~\citep{Gundacker_2020}. Each SPAD is a microscopic independent Avalanche PhotoDiodes (APDs),
operated in Geiger-mode, i.e., counting single photons. However, when combining thousands of microcells, the dynamic range
of a SiPM can go from a single to thousands of photons for each
square millimeter. The SiPM matrix wafer is prepared during the initial production and later wired and assembled. The thickness of the sensor is of the order of one millimeter, making the detector very compact. SiPMs are operated at bias voltage of few tens of volts, carefully set to provide optimal performance in terms of single photon response, linearity, and current noise. The small size also makes SiPMs insensitive to regular ambient magnetic fields.

\begin{figure}[h!t]
    \centering
    \includegraphics[width=1\linewidth]{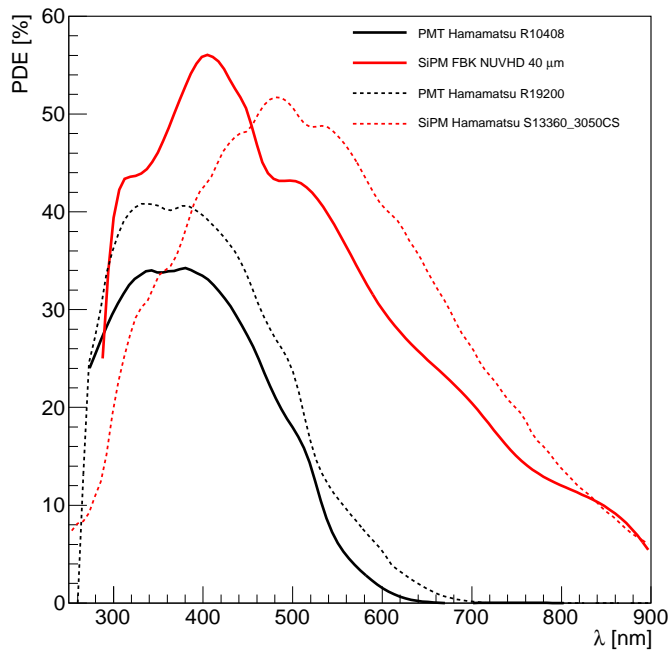}
    \caption{\label{fig:pde} PDE curves for the reference PMTs and SiPMs for this paper. The current MAGIC PMT is a Hamamatsu R10408 mounted in the MAGIC cameras (black solid curve~\cite{Nakajima:qe2013}).  
    The solid red curve displays the FBK NUVHD3 40~$\mu$ sensor \cite{gola}. These two are used in this work. For further comparison, the dotted red curve displays the SiPM S13360 from Hamamatsu ~\cite{Otte:2016aaw} and the dotted black curve reports the CTA LST-1 PMT of type Hamamatsu R19200~\cite{nakajima}.}
  \end{figure}
  
%The photon detection efficiency (PDE) of SiPMs, see \autoref{fig:pde}), is normally higher than that of photomultiplier tubes (PMTs) in the blue band and very much higher in the red band.
In \autoref{fig:pde}, we compare the photo detection efficiency (PDE)
of one benchmark PMT Hamamatsu R10408~\cite{Nakajima:qe2013}, currently installed in MAGIC, the PMTs Hamamatsu R19200~\cite{nakajima}, currently installed in the first CTA large-size telescope prototype (LST-1) and two benchmark SiPMs that are under scrutiny for the use in CTA cameras. The SiPMs are a commercial Hamamatsu S13360\footnote{Datasheet at \url{https://www.hamamatsu.com/resources/pdf/ssd/s13360\_series\_kapd1052e.pdf}}~\citep{Otte:2016aaw} and a NUVHD3 sensor by Fondazione Bruno Kessler (FBK)~\citep{gola,Depaoli:2023kgt}. As one can see, the SiPMs have a peak PDE in a part of the spectrum at significantly larger wavelengths than that of the PMTs. They also feature a good efficiency in the red part of the spectrum where PMTs are usually less sensitive. 

In the following, we take the FBK NUVHD3 photosensor as a case example. This is a precursor of a class of SiPMs under constant development. For example, the CTA small-size telescope prototype \citep{Adams:2022ynt} mounts version NUVHD6 of this sensor, however with performance not dissimilar to the NUVHD3 considered here~\citep{Depaoli:2023kgt}. The SiPM development is mostly focused on the reduction of dark noise and optical cross-talk.  

One significant source of noise for SiPMs is the optical crosstalk generated by secondary photons created during the charge recombination. Such spurious photons may trigger secondary breakdowns in neighboring cells~\citep{Rech:2008}. The effect is a randomized enhancement of the amplitude of the measured signal. 
This effect can be reduced with specific design geometry such as building trenches among neighboring cells~\cite{Depaoli:2023kgt}. Also the use of light-guides in front of SiPM allows to reduce this effect because spurious photons are not reflected back to the sensor but are free to escape outward~\citep{Rehbein:2020gxm}. The irreducible optical cross-talk must be modeled during signal reconstruction. However, the SiPMs main drawback with respect to PMTs remains their reduced size, which cannot exceed 1-2 cm$^2$ with current technologies, due to the large capacitance generated by the compact geometry. In case a large sensitive area must be covered (as that of the focal plane of an IACT, of the order of 2~m$^2$), particular solutions must be adopted. These are: a) the positioning of optical funnels, in the form of either hollow or solid concentrator\footnote{The classical shape of such funnel is the compound parabolic concentrator described in \cite{winston_cone,winstoncone}} at the entrance of the individual SiPMs, that increase the sensitive area by guiding the light to the photosensor surface, e.g. as done by \citet{Huber:2011sui,Aguilar:2014oba} or b) the combination of a number of photosensors into one larger pixel via summation, possibly also combined with a light funnel~\cite{Rando:2015jpa,Hahn:2019mek}, to avoid a too large number of channels. 
%One of the successful attempts to combine SiPM matrices into arrays with performance similar to a single PMT is reported in \cite{Ambrosi:2016emj,Depaoli:2023kgt} where a signal pulse width of 1-2~ns is reported, very close to the performance of currently used PMTs.

An example of a telescope using solid-state photosensors is that of the First G-APD Cherenkov Telescope (FACT)~\citep{Biland:2014fqa}. The FACT telescope is a single-dish 4~m-diameter telescope located at the MAGIC site, built upgrading a decommissioned HEGRA telescope and successfully operative for about a decade. FACT has
successfully operated for several years a Geiger-mode a\-va\-lan\-che
photodiode camera described by \citet{Biland:2014fqa}. G-APD photosensors allows FACT to routinely operate even under strong moon level \citep{FACT:2013irf} allowing for a duty cycle increase of 30\%~\citep{Dorner:2021mik}. 
MAGIC is currently hosting four SiPM clusters at the edges of the camera. They are only test installations not used for regular data analysis. One of the clusters hosts seven and the other nine $6\times6$~mm$^2$ photosensors per pixel. These are read out with the standard MAGIC data acquisition system~\cite{Rando:2015jpa,Hahn:2019mek,2018NIMPA.912..259H}. 
%The CTA project \cite{cta-site} is foreseeing all its small-size telescopes (SSTs) to be equipped with SiPMs \cite{Aguilar:2014oba,Heller2017}. As for the CTA LSTs, the use of SiPMs is also considered as a valid alternative to PMTs to improve their performance. 
% The goal of this work is to evaluate the performance of a SiPM-based upgrade of a MAGIC-like camera. For this purpose, we are mostly interested in the wavelength dependence of the PDE. The detailed technical implementation of a specific solution for such an upgrade is beyond the scope of this paper. Nonetheless, we will come back on the feasibility of a SiPM upgrade in \autoref{sec:discussion}. 

%\medskip
%The characteristics of the photosensors is not the only element that affects the performance of the instrument. Atmospheric absorption of the Cherenkov radiation also alter its spectrum significantly as will be discussed in \autoref{sec:yield}. The light yield is affected by several factors such as the mirror reflectivity, the transmission of the camera window, besides the PDE. Furthermore, the Cherenkov spectrum itself generated in EASs changes in function of the energy and incident direction of the primary particle inducing such shower. The full spectral characterisation is the subject of the next section.

Even while we do not aim at a full simulation of a specific hardware solutions, in order to make the assessment of performance realistic, we adjusted a number of parameters related to SiPMs in the Monte Carlo simulation and analysis. This is presented in \autoref{sec:simulation}.

%%%%%%%%%%%%%%%%%%%%%%%%%%%%%%%%%%%%%%%%%%%%%%%%%%%%%%%%%%%
%
%
% SEC 3 LIGHT yield AND SPECTRUM
%
%
%%%%%%%%%%%%%%%%%%%%%%%%%%%%%%%%%%%%%%%%%%%
%\newpage
\section{Differential light yield at the photosensors}
\label{sec:yield3}
Considering the difference in PDE between PMTs and SiPMs is wavelength-dependent, it is important to also spectrally characterise the light impinging the photosensors. There are several effects that affect the spectrum hitting the camera pixels: first, the Cherenkov spectrum at ground, that  may non trivially change in function of the intrinsic energy of the primary particle, the particle type, and  the altitude above the horizon (or the zenith angle ZA or zenith distance ZD). In principle, different atmospheric conditions can affect as well the Cherenkov light extinction in a wavelength-dependent way. This effect is less general to model and it is less severe. 
Other factors are the reflectivity of the mirror facets, the transmissivity of the eventual entrance window (in the case of MAGIC made of a plexiglass sheet, see later), and the spectral properties of the light concentrators. All these factors contribute to shaping the wavelength-dependent light yield at an IACT camera. 

\subsection{Cherenkov light yield from gamma-ray and cosmic-ray induced atmospheric showers}
\label{subsec:cherenkov}

The Cherenkov spectra at ground studied for this work are
obtained from dedicated Monte Carlo simulations. These are based on a
customised version of the publicly available software for simulations of air showers, dubbed Corsika (COsmic Ray
SImulations for KASCADE), which is used within the MAGIC collaboration~\citep{Heck:1998vt}. To model the light extinction in the atmosphere, we
used an atmospheric model specifically 
produced for the sky at the ORM~\citep{Haffke:2007}.
%, with a molecular profile obtained with the NRLMSISE-00 global model~\citep{Picone:2002}, slightly changing over the year, plus an aerosol distribution obtained with the Elterman profile\footnote{La Palma is the site that hosts the MAGIC   telescopes, and is the chosen site for the   northern array of CTA. The atmospheric model used here is currently   obsolete, and today preciser estimations of atmospheric extinction are obtained from a cross reference between the GDAS model and  IDAR-based information~\citep{Gaug:2014hja}. We do not expect strong significant differences with different models, however, the comparison of different profiles is out of the scope of this work.} 
We simulated 180,000 gamma ray events and 60,000 proton events in total, at different impact parameters to simulate a realistic condition. To evaluate the effect of the sky position and primary energy on the impinging Cherenkov spectrum, we considered three incident angles ($5, 45, 65\degree$) and three energies for gamma rays: $60, 400, 1500$~GeV and an energy of $180$~GeV for protons\footnote{Considering protons produce copious pions in hadronic interactions, electromagnetic subshowers are initiated by a charged particle with roughly one third of the primary proton energy, hence the factor of three in energy in comparison between primary gamma rays and protons}. This  corresponds to 20,000 simulated events per set.  

At the emission point, the Cherenkov yield is generated with a $\lambda^{-2}$ spectrum~\cite{Rossi:1941}. This spectrum is attenuated with strong chromaticity by the atmosphere molecular content. At short wavelengths, ozone plays the major role. At longer wavelengths, Rayleigh scattering with Nitrogen and Oxygen becomes the strongest attenuating agent. IACTs are typically located in very clean and dry atmospheric environments, whereas the conditions are practically those of free troposphere, with the aerosol content playing a significant role only
in non-optimal atmospheric conditions \cite{fruck:2022}. 

\begin{figure*}[h!t]
    \centering
    \includegraphics[width=0.24\linewidth]{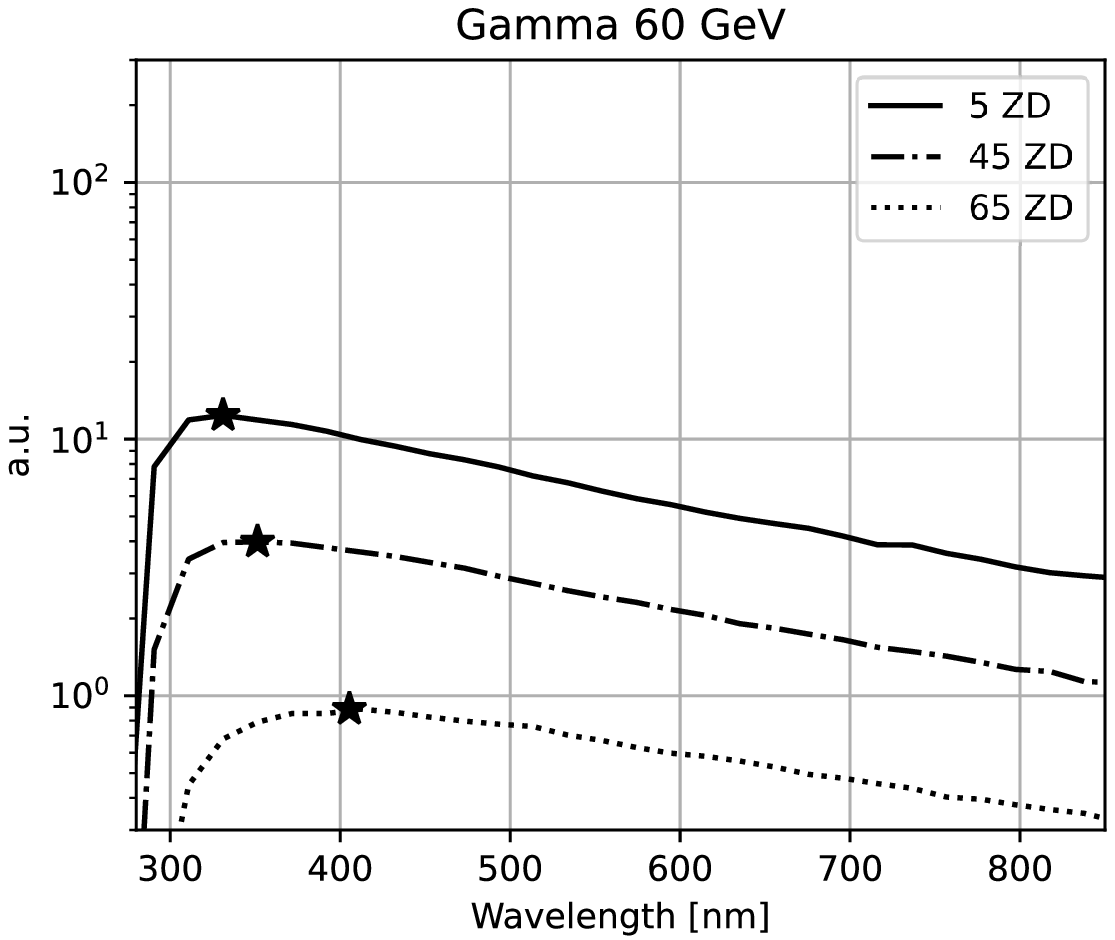}
    \includegraphics[width=0.24\linewidth]{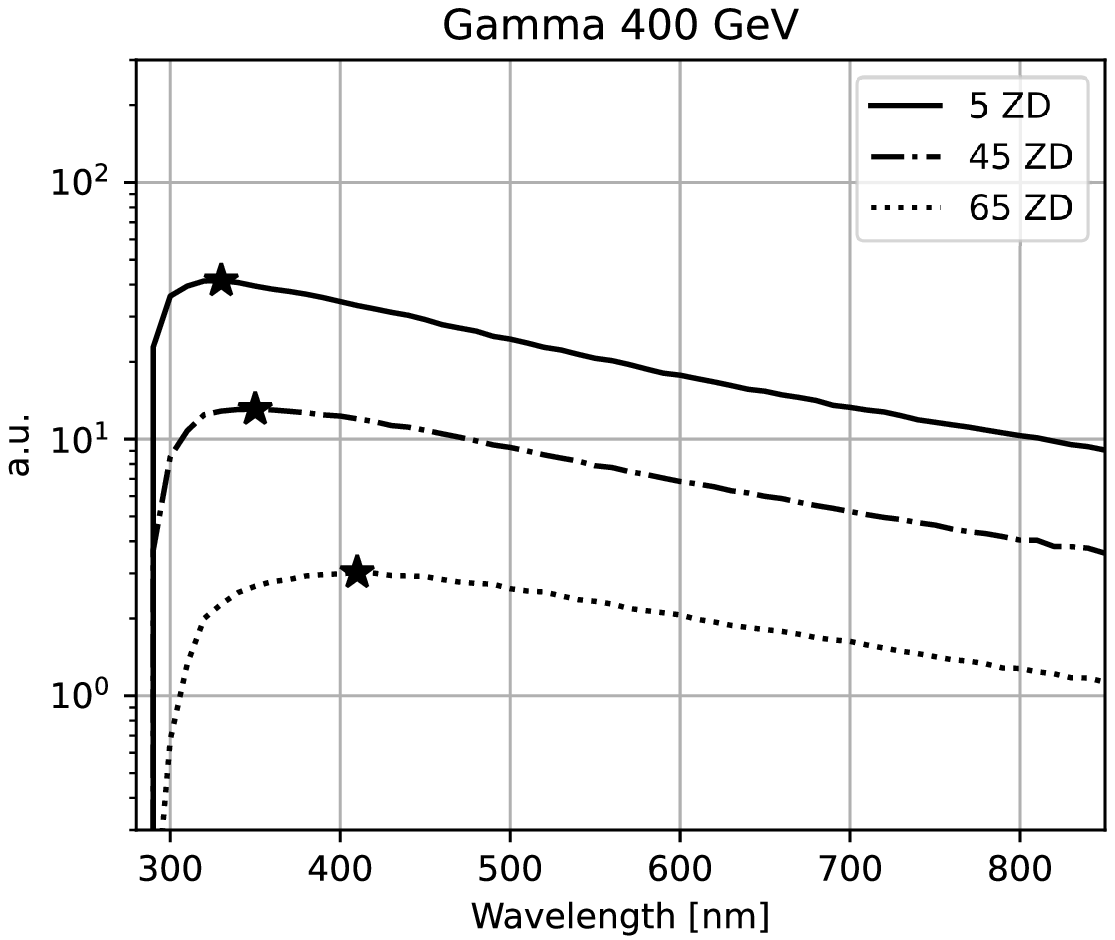}
    \includegraphics[width=0.24\linewidth]{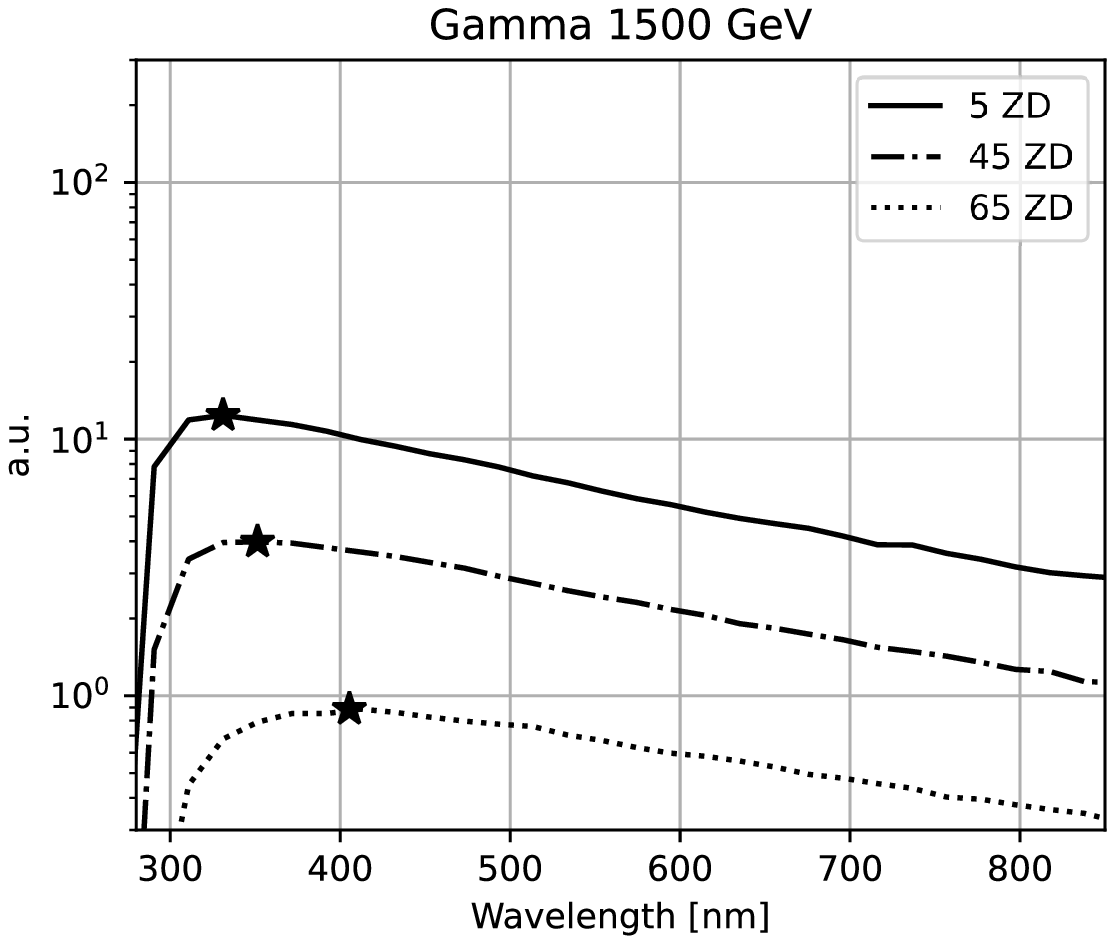}
    \includegraphics[width=0.24\linewidth]{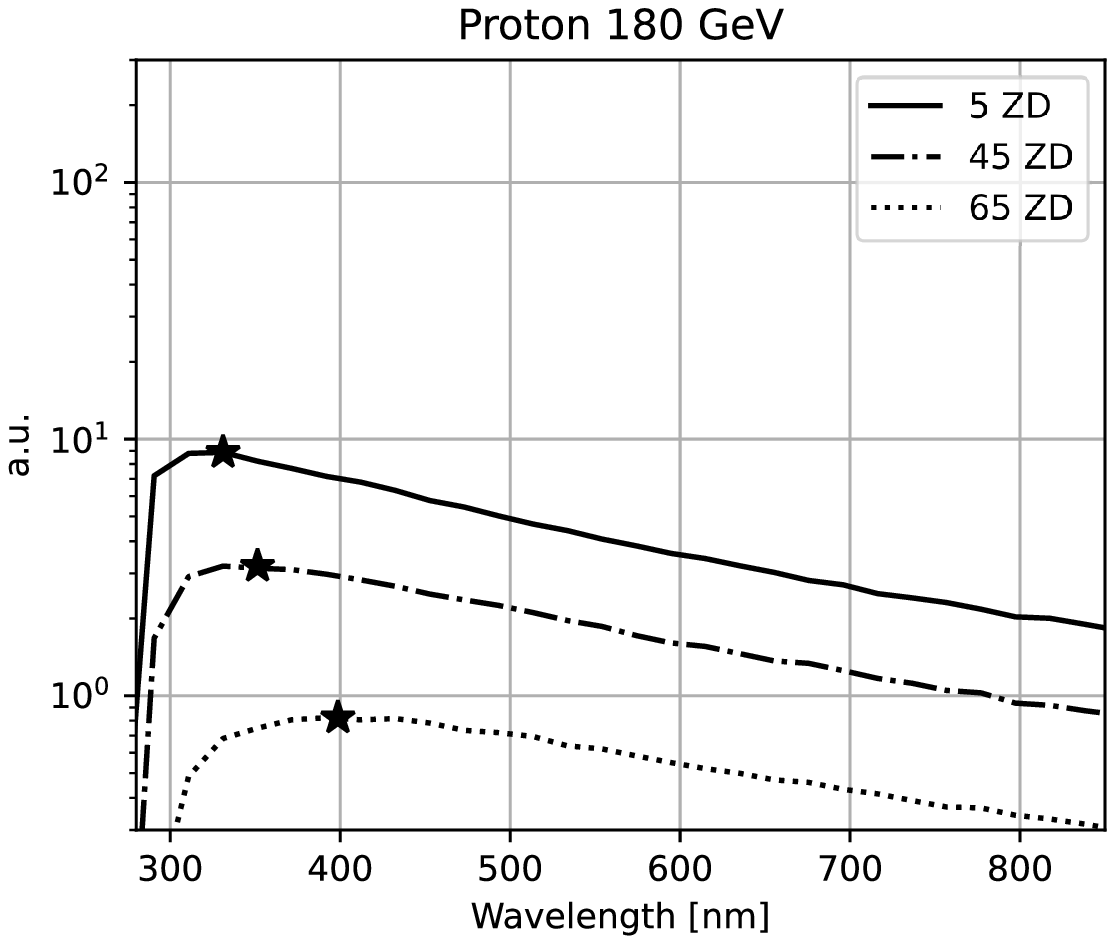}
\includegraphics[width=0.24\linewidth]{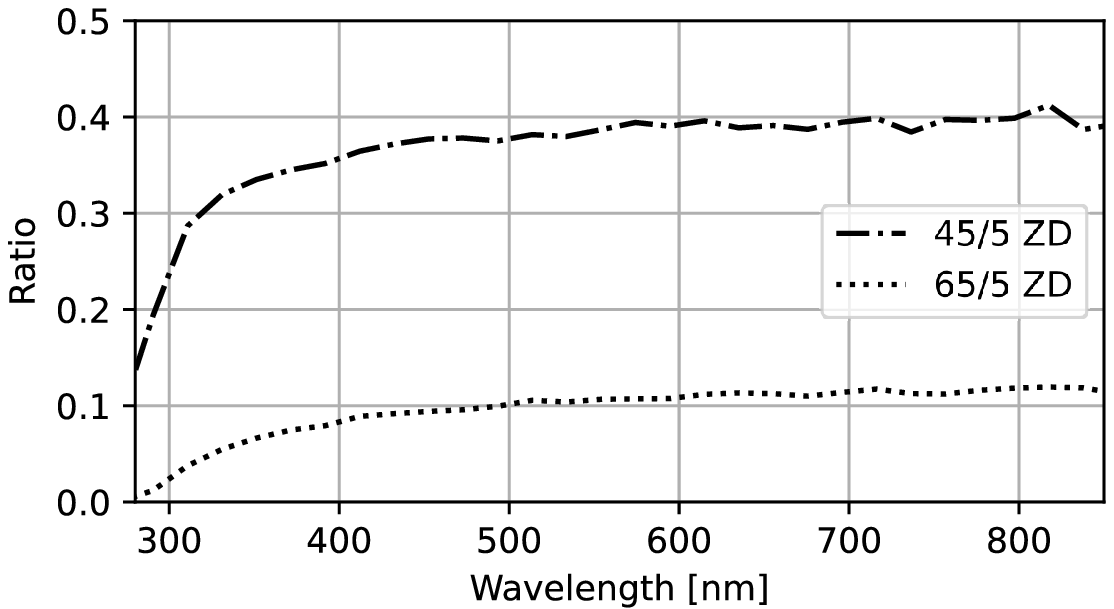}
\includegraphics[width=0.24\linewidth]{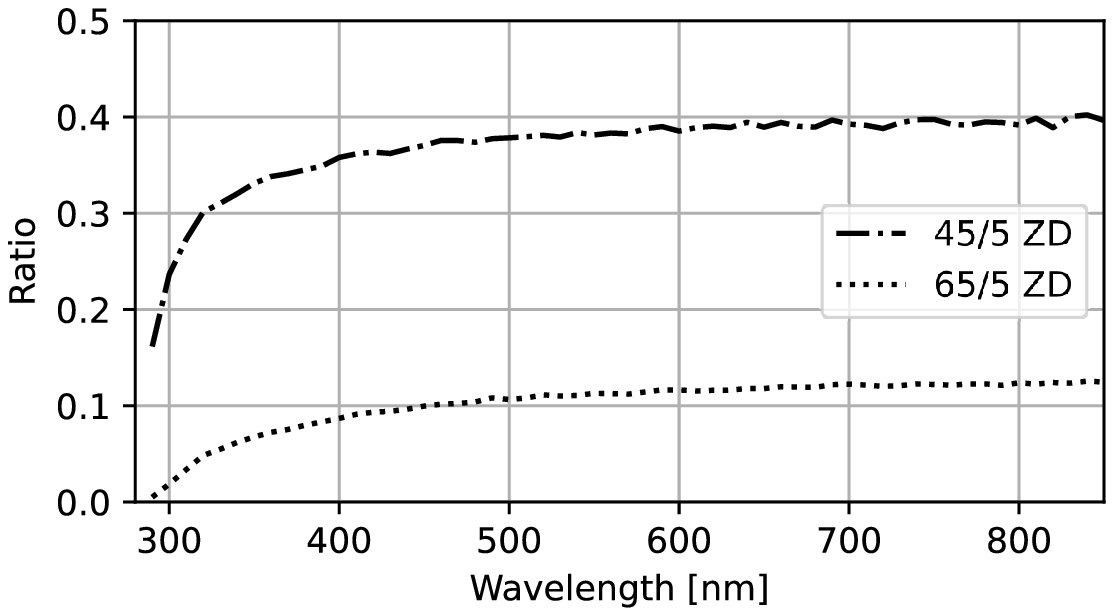}
\includegraphics[width=0.24\linewidth]{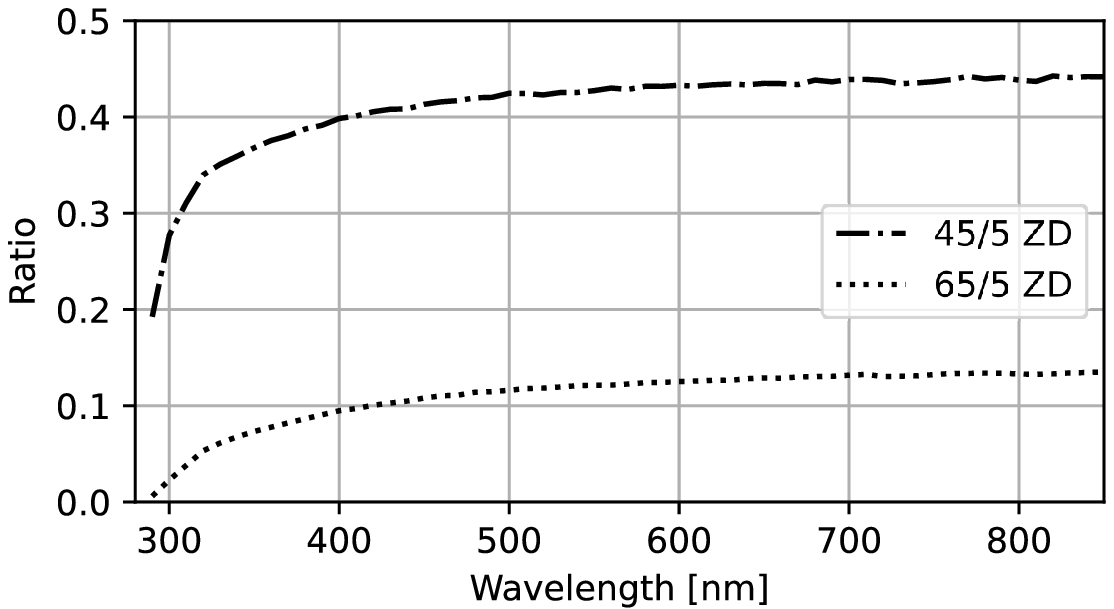}
\includegraphics[width=0.24\linewidth]{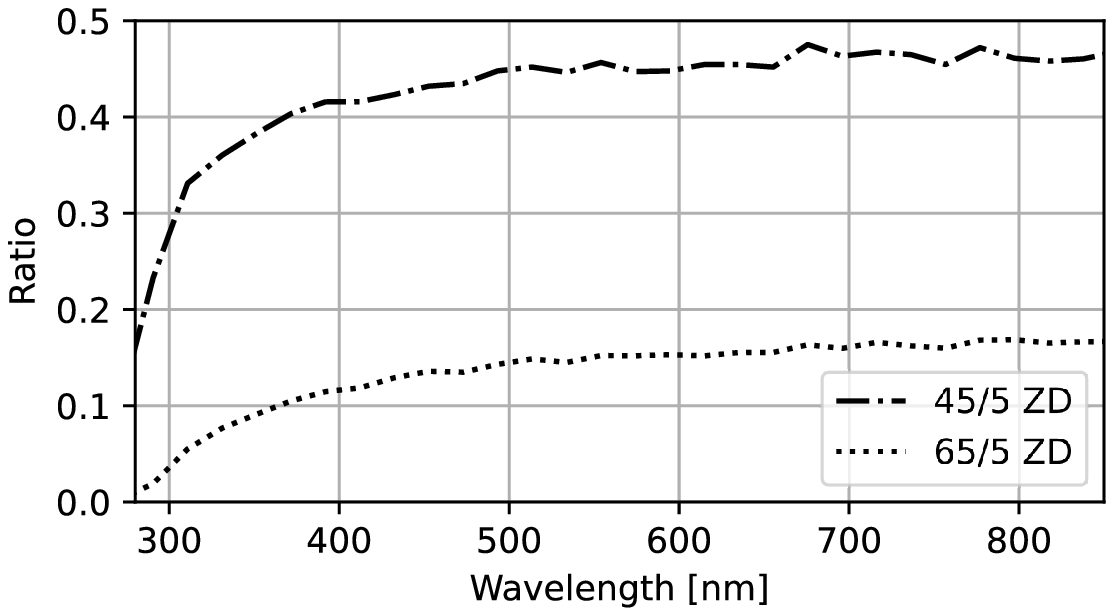}
    \caption{\label{fig:ch_spectrum}Cherenkov spectrum at the focal plane, in
      function of the energy and incoming direction of incident gamma rays (three leftmost columns) and incident protons (rightmost columns), considering only atmospheric absorption (e.g. not the mirror reflectivity loss). The top row reports the
      total number of photons in arbitrary units in a the standard MAGIC cameras for three
      different energies for primary gamma rays: 60 GeV (left), 400 GeV (centre-left) and 1500 GeV
      (center-right) and for the energy of 180~GeV for protons. In each pad, three incident zenith angles are reported: $5^\circ$
      (solid line), $45^\circ$ (long-dashed line) and $65^\circ$ (short-dashed line). The bottom row shows the ratio of
      the individual spectra simulated for higher zenith angles with respect to the 5$\degree$ case. The black stars on the plots indicate the peak of Cherenkov spectra. The
      corresponding numerical values are reported in \autoref{tab:ch_peak}. 
      }
  \end{figure*}
  
\autoref{fig:ch_spectrum} shows the simulated Cherenkov spectra at the focal plane for primary gamma rays and protons at different energies and zenith angles. The spectrum is in arbitrary units and consider only the atmospheric absorption. No other wavelength-dependent effects, e.g. the mirror reflectivity, are considered. The curves are obtained by summing over all the events in a batch that give a detectable number of photons. This varies significantly event by event even with the same settings, due to the stochasticity of the EAS development and the simulation. For each spectrum, the peak position is identified as the average of the three highest points of the curve\footnote{For this reason, the statistical uncertainty on the peak position cannot be computed. Instead, the sample was split into two subsets to check the reproducibility of the peak position. For simulations at 5 and 45 deg zenith angle we found no difference. Only in the case of simulations with a zenith angle of 65 deg, the peak position was shifted by a few nm, probably due to the flatter maximum in the spectra.}. The peak positions are reported in \autoref{tab:ch_peak}. 

\begin{table}[ht]
  \centering
  \begin{tabular}{l|ccc}
    \hline
    \rowcolor[gray]{.8} \multicolumn{4}{l}{Cherenkov spectra peak wavelength} \\
    \hline
    & 5$\degree$ & 45$\degree$ & 65$\degree$ \\
    \hline
    $\gamma$ 60 GeV & 331 nm & 351 nm & 405 nm \\
    $\gamma$ 400 GeV & 330  nm & 350 nm & 410 nm \\
    $\gamma$ 1500 GeV & 330 nm & 340 nm & 403 nm \\
    \hline
    $p$ 180 GeV & 331 nm & 351 nm & 398 nm \\
    \hline
    %\rowcolor[gray]{.8}    \multicolumn{4}{l}{Relative shift w.r.t. increasing inclination}\\
    %\hline
    %& 5$\degree$ & 45$\degree$ & 65$\degree$ \\
    %\hline
    %60 GeV   & --& +5\% & +22\% \\
    %\hline
    %400 GeV  & --& +5\% & +25\% \\
    %\hline
    %1500 GeV & --& +6\% & +26\% \\
    %\hline
    %\rowcolor[gray]{.8}    \multicolumn{4}{l}{Relative shift w.r.t. increasing energy}\\
    %\hline
    %& 5$\degree$ & 45$\degree$ & 65$\degree$ \\
    %\hline
    %60 GeV   & \multicolumn{1}{c|}{--}& \multicolumn{1}{c|}{--}& \multicolumn{1}{c}{--}\\
    %400 GeV  & \multicolumn{1}{c|}{-1\%}& \multicolumn{1}{c|}{-2\%}& \multicolumn{1}{c}{0\%}\\
    %1500 GeV & \multicolumn{1}{c|}{-2\%}& \multicolumn{1}{c|}{-2\%}& \multicolumn{1}{c}{0\%}\\
    %\hline
    %\rowcolor[gray]{.8}    \multicolumn{4}{l}{Cherenkov photon ratio 500 nm/350 nm}\\
    %\hline
    %& 5$\degree$ & 45$\degree$ & 65$\degree$ \\
    %\hline
    %60 GeV   & \multicolumn{1}{c|}{0.63}& \multicolumn{1}{c|}{0.73}& \multicolumn{1}{c}%{1.06}\\
    %400 GeV  & \multicolumn{1}{c|}{0.60}& \multicolumn{1}{c|}{0.72}& \multicolumn{1}{c}{1.02}\\
    %1500 GeV & \multicolumn{1}{c|}{0.61}& \multicolumn{1}{c|}{0.71}& \multicolumn{1}{c}{0.99}\\
    %\hline
  \end{tabular}
  \caption{\label{tab:ch_peak} 
  %(top block) 
  Position of the maximum of the
    Cherenkov spectrum at ground as a function of the energy (rows) and
    incident angle of the primary gamma ray (columns). 
    %(second block) Relative redshift of the maximum emission for increasing inclination of the incoming gamma ray with respect to the vertical case. (third block) Relative blueshift of the maximum emission for increasing     energy of the gamma ray with respect to the 60-GeV case. (fourth block) Ratio of     the number of Cherenkov photons emitted at 500 nm and 350 nm, respectively.
    Statistical uncertainties cannot be accurately computed as discussed in the text and are not reported.
    }
\end{table}

From \autoref{fig:ch_spectrum} one can see that with increasing incident angle of the primary particle, the absolute photon yield from EASs decreases significantly. This is due to the increased air mass the Cherenkov light travels through before reaching the ground, which results in higher extinction.  A second effect is that the
  position of the maximum of the Cherenkov spectrum moves to longer wavelengths
  (redshift) with increasing incident angle of the gamma ray. This is due to the
  differential extinction of the atmosphere at different
  wavelengths (Rayleigh scattering). The maximum becomes significantly redshifted with increasing incident angle,
  up to 20\% between $5^\circ$ and $65^\circ$, but almost independently of the energy of the gamma ray. This behavior is especially interesting in the context of observation at large zenith angles (above $75^\circ$) which are currently being carried out in the field \citep[see, e.g.][]{MAGIC:2020xry}.
The insignificant energy dependence of the Cherenkov peak allows us to 
neglect any energy-dependent effect for the computation of
the photon yield in the remainder of this work. We will revisit the energy dependence during
the examination of the interplay with the night sky background (\autoref{subsec:nsb}) and the overall instrument sensitivity (\autoref{sec:sensitivity}).

It is also interesting to compare the Cherenkov spectrum emitted by electromagnetic showers and hadronic (sub)showers. In \autoref{fig:ch_spectrum} (rightmost column) we report the Cherenkov spectra from electromagnetic subshowers generated by 180~GeV protons. The trend of the maximum Cherenkov radiation shifting to longer wavelengths and decreasing with increasing inclination of the primary particle shows the same regularity between gamma ray and proton induced atmospheric showers.

\subsection{The light of the night sky}
\label{subsec:nsb}
IACTs operate in the presence of the optical background from the Light Of the Night Sky (LONS). The LONS is composed of starlight, zodiacal light and airglow. The LONS is strongly site dependent and also strongly vary with the time of the year, zenith angle and moon phases, among other things. For an IACT, it constitutes the so-called Night Sky Background (NSB).

\begin{figure}[h!t]
    \centering
         \includegraphics[width=1\linewidth]{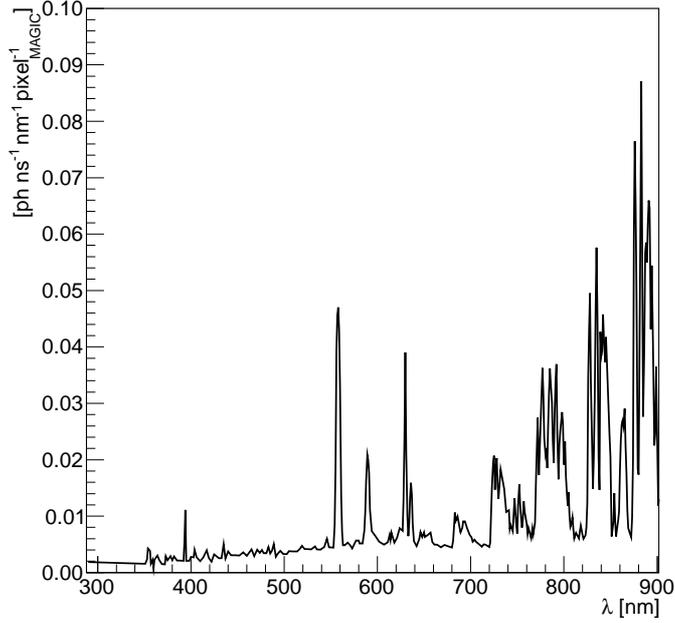}
    \caption{\label{fig:nsb_spectrum}The spectrum of the NSB at La Palma. The data are taken from~\citet{Benn:1998nsb} and converted to number of
      photons reaching one MAGIC pixel in one nanosecond. 
      %\jsc{hmm, now I do not understand. If you added/removed points than bin size is different for each point, so let's say sometimes it is 1 nm, sometimes 5 nm, and then if you give the flux per bin, it would be also in one place per 1nm and in another per 5nm. I guess you did not really do it like this, otherwise it would make the plot basically useless since you cannot compare relatively the Y values. Or do you mean that you started with some binning, and when you added removed points you just interpolated/averaged? In this case the Y axis is per original size of the bin (that can be quoted in the caption\ca{so I used the publicated plot and extracted the points from there, Obviously, I could not be that precise to do it each nm so sometimes, I used larger steps and then there is no interpolation here but simply a line added to the TGraph (this option however uses an interpolation if I am not mislead. As for the integrals: yes I used and interpolation and I explicitly set the wavelength range to the same and used bins of 1 nm for all calculations. I hope this helps?})}
      }
  \end{figure}

\autoref{fig:nsb_spectrum} shows a typical NSB at the ORM obtained from data from \citet{Benn:1998nsb} and converted for a pixel size of MAGIC.
%; in my opinion this is redundant as we explained the data in the figure caption}.\md{ONe reference in the text is not bad, people sometimes do not go directly to the figure.}
There are two major differences between the NSB and
Cherenkov spectrum: a) the NSB \textit{increases} with the wavelength and is concentrated at the red and infrared part of the spectrum, and b) the NSB shows several very prominent \textit{peaks}
over the steadily increasing background. This is directly related to the
emission spectrum of the air-glow. In reality, this is an average picture, as discussed above. 

Having the pixels of an IACT camera hit by spurious photons creates two problems: a) fake triggers up to saturation of the data acquisition system and b) noise within the
signal with a consequent deterioration of the signal-to-noise ratio (SNR). We have already discussed that SiPMs are more sensitive than PMTs in the red part of the spectrum (see \autoref{fig:pde}). It will be therefore important to weigh this effect properly.

\subsection{Other optical elements affecting the light yield}
A number of optical elements of IACTs exhibit wavelength-dependent behaviour, in particular the mirror dish facets, the light concentrators attached to the camera pixels and the plexiglass entrance window. 
The switch to a SiPM-based camera would not require modifications of these aspects but nevertheless, their wavelength-dependency still needs to be considered in this study, as the optics might affect the relevance of the PDE of the SiPM at different wavelengths.
We pay special attention to the camera entrance protection window (see \autoref{subsec:window}), whose wavelength-dependent transmission can provide a cost-effective way to improve the performance of a SiPM-based camera, as we will see later. 

\paragraph{Mirrors} The mirrors of the MAGIC telescopes  (see \cite{Doro:2008zz,Doro:2009xqa,will:2019} have a specific spectral reflectivity, that depends on the material used for the mirror surface.
Their reflectance is achieved by an aluminium alloy, varying between different mirror designs, 
%\ca{maybe some references?}
%\md{no need IMHO}
coated with a protective quartz layer. The thickness of this layer affects the position of the maximum spectral reflectivity. 
\autoref{fig:window} shows the spectral reflectivity of a typical MAGIC mirror \citep{Bastieri:2005pq}. 

\paragraph{Light funnels} The cathodes of the photosensors in the current MAGIC cameras are hemispherical, causing some dead area between neighbouring pixels. Light concentrators, often called ``Winston cones'' \citep{winston_cone,winstoncone, Henault:2013dsx}, reduce the size of this area. At the same time they reduce the number of NSB photons entering at large field angles, increasing the SNR for the faint gamma-induced signal. 
In MAGIC, the light guide is made of a few cm high ultra-thin metallized foil shaped into a parabolic geometry with a larger hexagonal entrance and a smaller circular exit, which is directly attached to the cathode of the PMT. 
The light concentrators are made from a plastic material with aluminised Mylar foil of about 95\% reflectivity above 300~nm~\citep{Okumura:2017qvd}.  \autoref{fig:window} shows the spectral reflectivity of this foil for light incident at 30$\degree$\footnote{The reflectivity depends slightly on the inclination angle.} This contribution is neglected for the light yield comparison study in \autoref{sec:yield4}, but it is fully considered for the study of performance in \autoref{sec:simulation} and \autoref{sec:sensitivity}.

\subsubsection{Camera protection window}
\label{subsec:window}

\begin{figure}[h!t]
    \centering
    \includegraphics[width=1\linewidth]{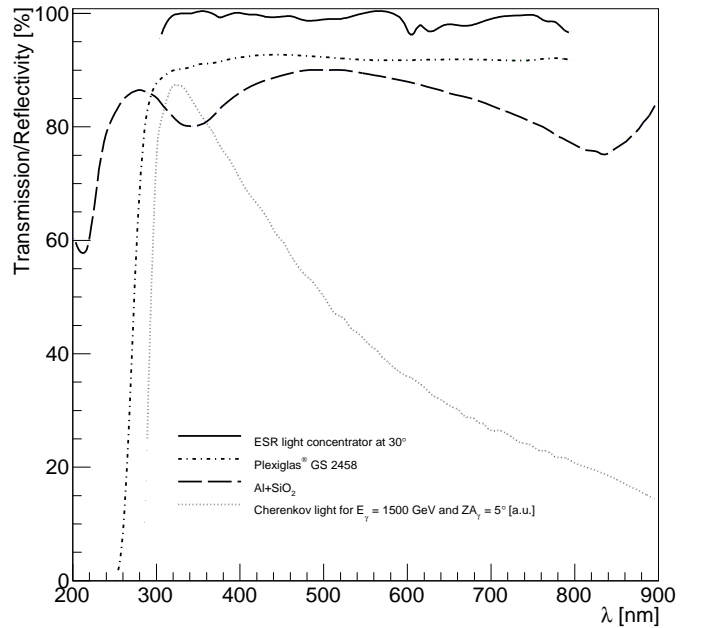}
    \caption{\label{fig:window} Transmission of the PMMA of type
      Plexiglas\textregistered GS Clear 2458 SC (dashed-dotted black line)
      and spectral reflectivity of the enhanced specular-reflector light concentrator foil~\citep{Okumura:2017qvd} at
      30$\degree$ incidence (solid black line) and of the quartz-coated all-aluminium MAGIC mirrors (long-dashed, \citep{Bastieri:2005pq}). The Cherenkov spectrum
      produced by a 1500-GeV gamma ray with 5$\degree$ incident angle is shown for reference (dotted gray line).}
  \end{figure}

  To protect the pixels and electronic from environmental conditions, the focal plane instrumentation of MAGIC is hosted in a thermally regulated, water- and light-tight container. 
  A transparent plexiglass window (or similar polymethyl methacrylate-like materials; PMMA) is used to seal it, while proving good light transmission, specially close to the maximum of the Che\-ren\-kov spectrum. As a further protection, the plexiglass window and the camera is closed by light- and water-insulating lids, closed during the day.
  %The focal plane instrumentation of MAGIC is hosted in a water- and light-tight container, which is also thermally regulated. 
  %When the telescope is performing observation, two lids open and the photosensors are exposed to the sky. 
%
   %The focal plane instrumentation of MAGIC is hosted in a water- and light-tight container, which is also thermally regulated. 
  %When the telescope is performing observation, two lids open and the photosensors are exposed to the sky. 
  %To protect the pixels from fast raising humidity or rapidly incoming clouds or rain, and to protect the pixels and the inside electronics from dust, a transparent plexiglass window (plexiglass, or similar materials like PMMA) is used, which show lightweight, mechanical resistance, reduced ageing and good transmission, specially close to the Cherenkov spectral peak. 
  The plexiglass window in front of the MAGIC camera is of kind
Plexiglas\textregistered~GS Clear 2458 SC. It has a transmission of
92\% at all wavelengths above 355~nm, which is slowly decreasing to 88\% at 
300~nm with a sharp cut-off at 280 nm (see \autoref{fig:window}).
As there is no sizable  
effect on the wavelength range of interest, as such, we can exclude this contribution for the spectral-response comparison between PMT and SiPM for the light yield comparison study in \autoref{sec:yield4} but this contribution is considered when performing the full telescope simulation of \autoref{sec:simulation} and \autoref{sec:sensitivity}.

%\subsubsection{Light Concentrators}
%\label{subsec:funnels}
%
%The cathodes of the photosensors in the current MAGIC telescopes
%cameras are hemispherical. This means that the focal plane area shows
%some dead-space among the photosensors. Light concentrators reduce the
%size of the dead area between the photon detectors and reduce the number
%of direct and indirect NSB photons entering at large
%field angles, increasing the signal-to-noise ratio for faint Cherenkov
%photons. Such light concentrators, often called “Winston
%cones” \citep{winston_cone,winstoncone} are commonly used~\citep{Henault:2013dsx}.
%%
%The light guide is made of an ultra-thin
%metallic foil with a circular exit face (directly attached to the cathode of
%the PMT) and an entrance face of hexagonal shape. The height of these
%lightguides is typically of few centimeters, and they are
%pre-assembled together with the pixel and later on mounted onto the
%camera. In the case of MAGIC, the light concentrators are made from a
%plastic material with aluminized Mylar foil of about 85\%
%reflectivity. The reflectivity curve slightly depends
%on the inclination angle. This fact is not taken into account into our simulations. For 30$\degree$, the reflectivity curve is shown in
%\autoref{fig:window}. 
%Again, the fact that the reflectivity for the light concentrators is
%totally independent on the wavelength in our range of interest, allows
%us to exclude them from the subsequent analysis. 

%%%%%%%%%%%%%%%%%%%%%%%%%%%%%%%%%%%%%%%%%%%%%%%%%%%%%%%%%%%
%
%
% SEC 4 LIGHT yield
%
%
%%%%%%%%%%%%%%%%%%%%%%%%%%%%%%%%%%%%%%%%%%%%
%\newpage
\section{Light yield with a SiPM-based camera}
\label{sec:yield4}
We now move to combine all the spectral-dependent factors described in the previous section to understand how these affect the detection of Cherenkov light from primary gamma-ray induced EASs as well as the noise by the NSB. We do this in \autoref{subsec:yield_photons}. Given the effect of the sensitivity of the SiPM to the red band and therefore to the NSB, we discuss a possible alternative technical solution to counteract this effect in \autoref{subsec:dichroic}.  

\begin{figure*}[h!t]
  \centering
  \includegraphics[width=0.49\linewidth]{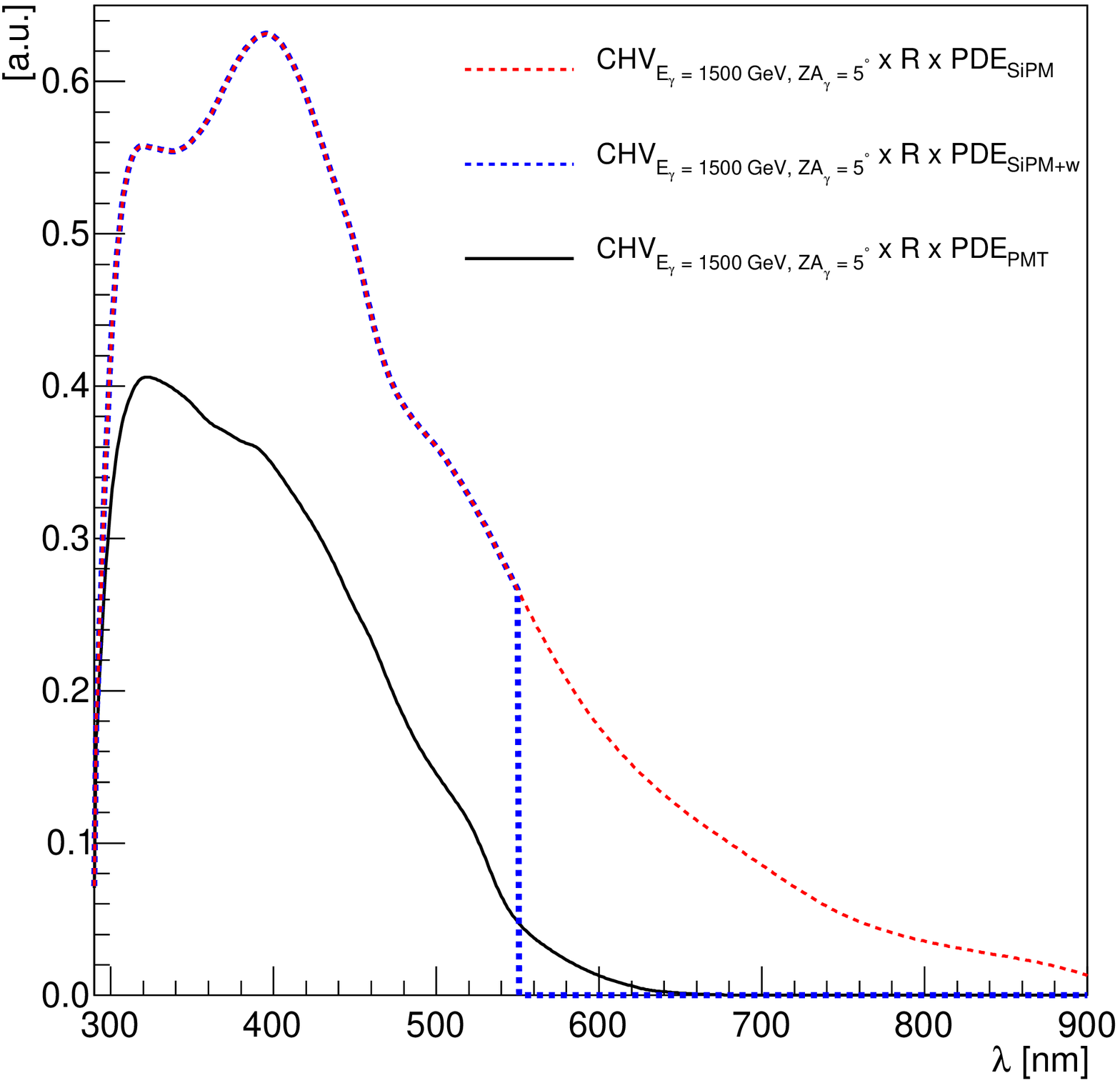}
  \includegraphics[width=0.49\linewidth]{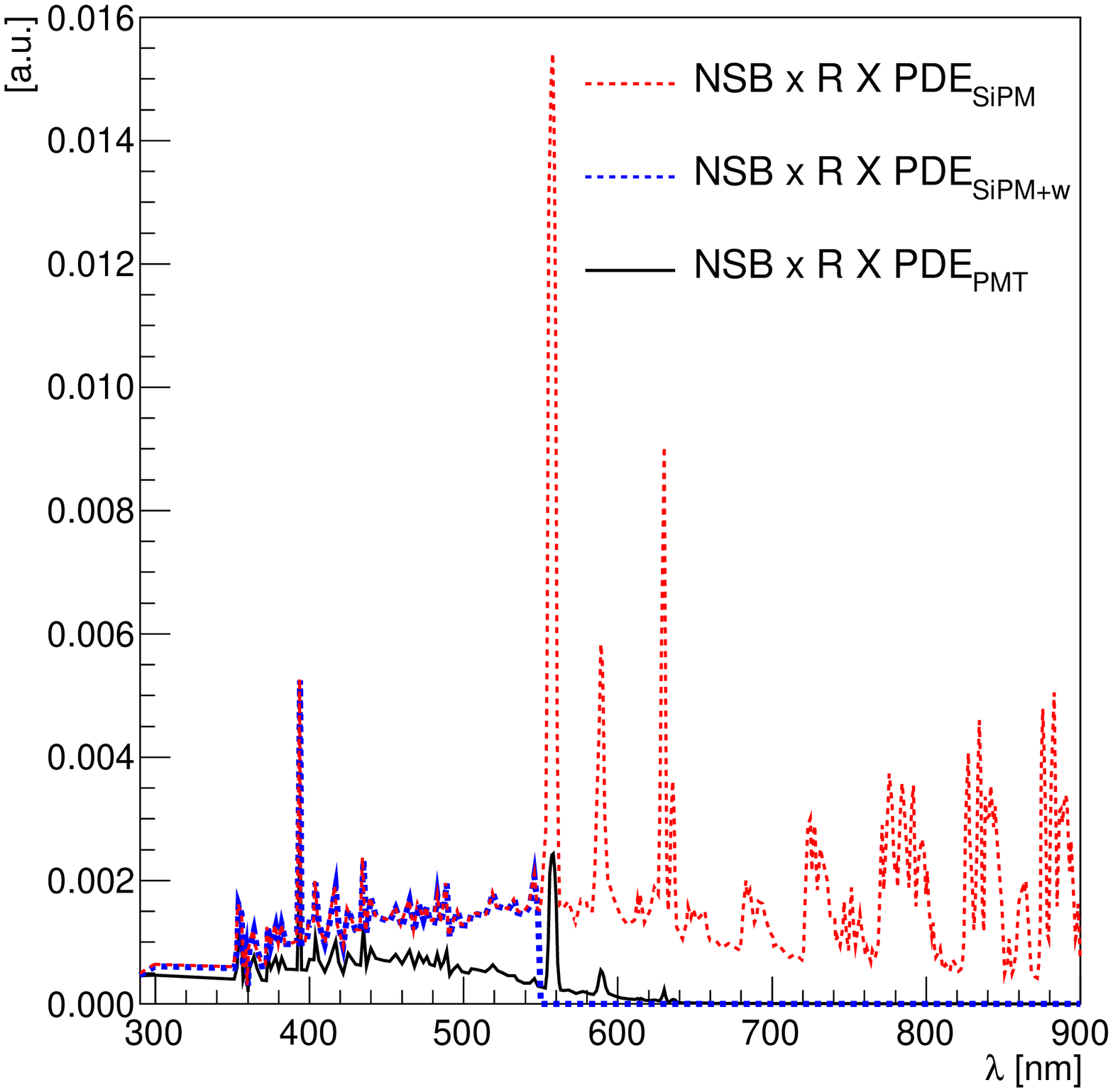}
    \caption{\label{fig:compare_numphot} (left) Cherenkov photons that
      survive the reflection onto the reflectors of the MAGIC telescopes and are subsequently detected by a PMT (black solid line) and a SiPM (red dashed line), respectively, for a primary gamma ray of 1500 GeV and 5$^\circ$ incident angle. (right) NSB spectrum weighed with the mirror
      reflectivity and PDE of a PMT (black solid line)
      and a SiPM (red solid line), respectively. The blue dotted line is computed by artifically filtering out all wavelengths above 550~nm for reasons discussed in \autoref{subsec:dichroic}.}
  \end{figure*}
  
\subsection{Cherenkov light and NSB yield}
\label{subsec:yield_photons}
\autoref{fig:compare_numphot} (left)
shows the Cherenkov signal yield that
survives the reflection onto the mirror dish of the MAGIC telescope and
the subsequent detection in a camera pixel, either consisting of a conventional PMT or the specific FBK SiPM considered in this study.
We limit our considerations to the case of a primary gamma ray of 1500 GeV, 
having shown that the amount of Cherenkov light at the ground simply scales
linearly with primary energy, but the spectrum is not changed (see \autoref{sec:yield4}). The wavelength-independent effects of the camera
window and the light guides are also here excluded. Already a quick inspection of the figures shows that the use of a SiPM strongly increases the signal in the blue part of the spectrum, allowing to gather more Cherenkov photons, but also makes the instrument more sensitive in the red part above 500~nm. A more quantitative comparison of \autoref{fig:compare_numphot}
is made in \autoref{tab:fold_numphot} (first rows, column "SiPM / PMT") in which we report the fractional increase of the integrals of the spectra displayed in \autoref{fig:compare_numphot} for the SiPM and PMT case\footnote{For reasons already expressed in \autoref{subsec:cherenkov}, we cannot compute the statistical uncertainties of these values.}. In all cases, the SiPM provides a significant boost in light yield with
respect to the PMT: the total photon yield increases from a factor $2.2$ to $2.5$ going from 5 to
65$\degree$ incident angle of the primary particle. This increase with the incident angle is due to the redshift of
the maximum of the Cherenkov spectrum as the Rayleigh absorption increases (see \autoref{subsec:cherenkov}).

\begin{table}[h!t]
  \centering
  \begin{tabular}{l|c|c}
    \hline
    \rowcolor[gray]{.8} 
    yield & SiPM / PMT & (SiPM+w) / PMT \\

    \hline
    \rowcolor[gray]{.9} 
        & \multicolumn{2}{c}{Cherenkov Light} \\
        \hline
    $\mathrm{ZA}_\gamma=5^\circ$ & 2.20& 1.76\\
    \hline
    $\mathrm{ZA}_\gamma=45^\circ$ & 2.28 & 1.76 \\ 
    \hline
    $\mathrm{ZA}_\gamma=65^\circ$ &2.52 & 1.84 \\
    \hline
     \rowcolor[gray]{.9}
       & \multicolumn{2}{c}{Night Sky Background} \\
    \hline
    & 5.30& 1.66\\
    \hline
  \end{tabular}
  \caption{\label{tab:fold_numphot} Fractional increase of Cherenkov photons produced by a primary
  gamma ray of 1500 GeV impinging at different zenith angles, as well as the increase of the NSB photons detected by a SiPM with
  respect to a PMT. The ratio is also reported for a technical solution applied to suppress/eliminate the sensitivity of the SiPM to the red band of the spectrum, which is based on the use of a low-pass filter ("SiPM+w"). 
  }
\end{table}

\medskip
The detection efficiency of the Cherenkov signal is not the only factor influencing the sensitivity of an IACT, because of the interplay with the NSB (\autoref{subsec:nsb}). The effect of SiPM and PMT on the NSB light yield is also shown in \autoref{fig:compare_numphot} (right). A visual inspection clearly shows that the behavior in the case of PMTs and
SiPMs is rather different: while in the former case, above 650~nm, the
NSB does not contribute significantly to the background light level of each camera pixel, instead in the SiPM case, it
does. Second, the contribution of spectral peaks present in the NSB changes in the two cases: in
the PMT case mostly the 557.7~nm peak is detected, while in the SiPM case,
many other significant peaks of the NSB contribute to the background level in the camera pixels. Looking again at \autoref{tab:fold_numphot} (bottom row, column "SiPM / PMT"), we quantify the increased number of photons with a factor of $5.3$ when using SiPMs instead of PMTs. These results are in line with those reported in~\cite{2018NIMPA.912..259H}.

\medskip
We asked ourselves whether such an additional "red" contribution could be avoided. This is in principle achievable in different ways: a) it is possible to realize a primary mirror with dichroic properties. For example \citet{Forster:2013xgr} demonstrated that multilayer dielectric mirrors could achieve this goal. We discarded this option for MAGIC, as replacing the entire reflectors is hardly feasible cost-wise b) another possibility 
is that of replacing the camera window (see \autoref{subsec:window}) with a dichroic plexiglass cutting the light above a certain wavelength. This is the topic of the next \autoref{subsec:dichroic}, c) finally, in principle one could also operate at the photon detector context and e.g. reduce the efficiency or the transmittance in the red part by special coatings\footnote{To our knowledge, this was not tested. Also possibly, a dichroic coating would increase the optical cross-talk of red photons trapped between the SiPM and the resin.}. This technical solution is not investigated here further. All in all, the effect of the three solutions would be similar: the light yield in the red part would be reduced. 
The anticipated impact of such solutions on the fraction increase of photons is reported in \autoref{tab:fold_numphot}, column "SiPM+w / PMT". One can see how such filtering effect has a moderate impact on the Cherenkov light yield, but a stronger one on the NSB, with relative weights similar as in the case of PMTs.

\subsection{A technical solution to reduce the photon yield of the red light}
\label{subsec:dichroic}
From \autoref{fig:compare_numphot}, it clearly emerges that if one could somehow cut out the sensitivity of SiPMs to the red part of the spectrum, the background light level due the the NSB could be reduced. Such a possibility may be offered, e.g., by substituting the neutral plexiglass window (\autoref{subsec:window}) with a dichroic camera window that would filter the dominant red part of the NSB spectrum by reflection or absorption. The modification of the entrance of the camera of IACTs is not a novelty in the field. It is used routinely in MAGIC during strong moonlight observations where a second protective camera window is mounted in front of the existing one~\citep{MAGIC:2017zph}. A similar solution was developed in VERITAS~\citep{Griffin:2015zsa}.
We will discuss the technical feasibility of this solution in \autoref{sec:discussion}. Here we show how to optimize the cut-off wavelength $\lambda_{\rm cut}$.

%For this reason we have simulated an alternative configuration in which the SiPM-based camera is coupled with a dichroic camera window (see Sec.~\ref{subsec:window}) that transmit only wavelength below 550~nm. We will discuss later on the technical implementation of this solution and the reason of this cut wavelength.

\begin{figure}[h!t]
    \centering
    \includegraphics[width=0.99\linewidth]{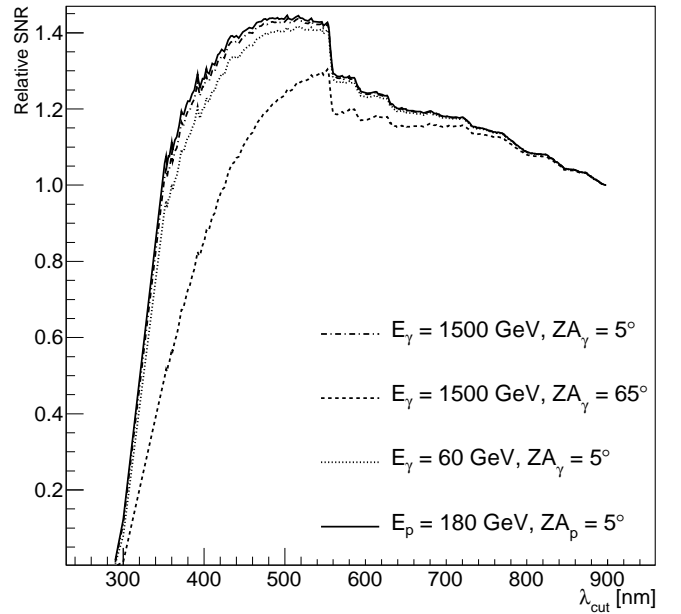}
    \caption{\label{fig:snr}Expected signal-to-noise ratio (SNR) of Cherenkov and NSB light detected with a SiPM-based camera using an ideal filter cutting wavelengths above $\lambda_{\rm cut}$ relative to the SNR obtained without such filter.
    Different curves correspond to different energies and incident angles of the primary gamma ray (see legend). For comparison the relative SNR is also shown for a proton-induced EAS (solid black line).
    }
  \end{figure}

To evaluate if the effect of the increased NSB level detected with SiPMs can be mitigated with a filter, we have performed a dedicated study taking into account the wavelength-dependent reflectivity of the MAGIC mirrors and the PDE of the NUVHD3 SiPM. We compare the amount of the predicted Cherenkov light from a gamma-induced air shower with the expected noise in the camera pixels from the NSB assuming that a perfect filter is used to reject photons with wavelength above $\lambda_{\rm cut}$. We compute the expected SNR$(\lambda_{\rm cut})$ as:
\begin{equation}
\label{eqn:snr}
\mathrm{SNR(\lambda_{\rm cut}})=\frac
%{\int_{\rm 290\,nm}^{\lambda_{\rm cut}}\frac{dN_{\rm Ch}}{d\lambda}T_{\rm atm}(\lambda) R_{\rm mir}(\lambda) {\rm PDE}_{\rm SiPM}(\lambda)d\lambda}
{\int_{\rm 290\,nm}^{\lambda_{\rm cut}}\frac{dN_{\rm Ch}}{d\lambda}(\lambda) R_{\rm mir}(\lambda) {\rm PDE}_{\rm SiPM}(\lambda)d\lambda}
{\sqrt{\int_{\rm 290 nm}^{\lambda_{\rm cut}}\frac{dN_{\rm NSB}}{d\lambda} R_{\rm mir}(\lambda) {\rm PDE}_{\rm SiPM}(\lambda)d\lambda}},
\end{equation}
%where $dN_{\rm Ch}/d\lambda$ is the simulated Cherenkov spectrum emitted in an air shower, 
where $dN_{\rm Ch}/d\lambda$ is the Cherenkov spectrum reaching the ground, 
%$T_{\rm atm}(\lambda)$ is the atmospheric absorption,
$R_{\rm mir}(\lambda)$ is the reflectivity of the mirror, 
${\rm PDE}_{\rm SiPM}(\lambda)$ is the PDE of the SiPM
and $dN_{\rm NSB}/d\lambda$ is the NSB spectrum observed at ground level.
The background component in the denominator is taken as a square root: due to AC coupling of the photodetectors they are only sensitive to the fluctuations of the light yield rather than its absolute level. 
We normalise ${\rm SNR}(\lambda_{\rm cut})$ to the SNR$(\lambda)$ without a filter, i.e., extending upper bound of the integral in \autoref{eqn:snr} to $\lambda = 900\,\rm{nm}$. The relative SNR is therefore ${\rm SNR}(\lambda_{\rm cut}) / {\rm SNR (900\,nm)}$ (see \autoref{fig:snr}).

\autoref{fig:snr} shows that the filter improves the relative SNR if $\lambda_{\rm cut}$ is at around 550\,nm. The improvement is of about 40\% for zenith angles of 5$^\circ$.  In case the observations are performed at a larger zenith angle, the gain in the relative SNR is lower, but still of order 30\%.
%\md{It is the value 1.3 at the peak for the 65 deg curve} \ca{I mean where do the 30 \% come from? I think for a non expert reader this consideration is a bit tricky to follow} \md{Let's try this and trust the reader!}

It is also interesting to note that the relative SNR pattern is similar also for protons of 180~GeV, which is in agreement with the results shown in \autoref{fig:ch_spectrum}. We display the light yield with the dichroic window as a dashed-blue line in \autoref{fig:compare_numphot}. We also show in \autoref{tab:fold_numphot} the relative gain regarding the photon yield considering such a technical solution. We see how the NSB now still increases with respect to the PMT case but from a factor of 5.30 to a factor of 1.66 with the modified camera window, while still keeping an increase of detected Cherenkov photons ranging from a factor of $1.76$ to $1.84$ depending on the incident angle of the primary gamma ray, as reported in \autoref{tab:fold_numphot}.

\bigskip
We remark that the above results cannot be directly used to estimate an increase in instrument sensitivity: the total photon
yield is not a proxy for the overall instrument performance. It is now mandatory to estimate the variation in sensitivity through dedicated Monte Carlo simulations, which we do in the next section.

%
%
%
% SEC 5 PERFORMANCE
%
%
%

\section{Simulation of a SiPM-based camera}
\label{sec:simulation}
\paragraph{Simulation of SiPM}
To fully compute the sensitivity with a SiPM camera,in addition to the different PDE, we also tuned the pulse shape, the shape of the single photo-electron (phe)\footnote{For simplicity we will also refer to the signal of a single triggered SiPM cell, as a ``photoelectron'', nomenclature commonly used for PMT signals.} distribution and trigger thresholds.  The early SiPM devices suffered from very broad pulses (see e.g. \cite{2012OptEn..51g4004Y}), the modern devices however have much narrower pulses (FWHM of 4.3~ns as reported in \citep{Depaoli:2023kgt}).
Even narrower 2--3~ns signals could be achieved, despite summation of signals from multiple individual SiPM devices \citep{Ambrosi:2016emj}. 
In our simulations we assume a 3~ns FWHM of a single phe pulse. With the additional assumed spread of individual phes, the effective pulse shape is 3.5~ns. 

Second, we also adjusted the response of the photodetector to a single registered phe. A PMT device has a rather broad distribution of a single phe charge. Moreover, during the multiplication of charge inside PMT, an atom might be ionized and accelerated against the flow of electrons. Such an ion hitting the photocatode produces afterpulses: large signals equivalent to even few tens of phes \citep{2007NIMPA.574..121A}. 
Due to their time delay, PMT afterpulses are not relevant for Cherenkov radiation. However, the afterpulses induced by NSB photons can mimic signals from small showers, limiting the performance of an IACT trigger. In contrast to this, the response of SiPM to light is completely different. 
While there are also delayed signals in SiPMs after a registered phe, they can be considered as pulses originating from not fully recovered cells. This makes their amplitudes smaller than a single phe~\citep{Otte:2016aaw}, which, combined with their relatively rare occurrence, makes them negligible effect for the SiPM performance in IACT applications.  

SiPMs have a single-phe resolution, i.e. the distribution of single phe amplitude is intrinsically narrow and clearly distinguishable from that of multiple phes. However, the effect of optical cross-talk introduces signal confusion. The probability of cross-talk depends on the SiPM device and its operation parameters. For example \citet{Otte:2016aaw} obtained values of 6\% -- 20\%. Multiple such signals can stack up, producing (with decreasing probability) multi-phe signals. This effect can affect both the Cherenkov light and NSB or dark rate induced phes. The difference of the single phe distribution for PMT and SiPM is therefore an important effect to consider. To take this into account, we produced a general single phe distribution representative of a SiPM, assuming a $\epsilon=15\%$ optical cross-talk. We use the formula of \citet{2010arXiv1006.3263G}, simplified to the case of a single true phe. The obtaining probability distribution of registered number $n$ of phes is: 
\begin{equation}
P(n) = (1-\epsilon)\times \epsilon^{n-1}.
\end{equation}
Furthermore we assume that each phe has an intrinsic amplitude spread of 10\%\footnote{The precise value of such a spread will depend on a specific hardware solution for amplification and registering the SiPM signals. We note however, that the number is expected to be small, and therefore its effect is expected to be negligible with respect to the optical cross-talk.}. 
Therefore, the amplitude of a $n$-phe signal is drawn from a Gaussian distribution with a standard deviation equal to $\sqrt{n} \times 10\%$ of its mean.

\paragraph{Tuning of Monte Carlo simulations and analysis}
The application of modified pulse width and single phe distribution for SiPM simulations requires additional changes in how those simulations are processed with respect to PMT camera simulations.
In order to integrate a similar fraction of the light pulse, in the case of SiPM we increase the size of the pulse integration window from 5 to 7 time slices. 
For the calibration of SiPM MC simulations we use the excess noise factor method, the same as for PMT cameras.
For the simulated $\epsilon=15\%$ case, the excess-noise factor ($F^2=1.16$) is comparable to the PMT one of $1.23$.

When a different photosensor is considered, as the SiPM considered in this work, the higher NSB level can affect the observations in two ways: increased trigger threshold (expressed in phes) and  higher noise in the images that could additionally require higher image cleaning\footnote{The image cleaning is the reconstruction procedure in which image pixels in the camera are selected by their signal level and their geometry.} thresholds. 
An Alternate Current (AC) coupling of the MAGIC camera pixels signals is installed to prevent the mean level of the NSB to affect the observations directly. 
Therefore the noise effect is only related to fluctuations of the NSB, which increases only as the square root of the total increase of the NSB, see also \autoref{subsec:dichroic}. 
%The stereo trigger of MAGIC is based on coincidence in time of three next-neighbour (3NN) pixels in each of the two telescopes. %
%Therefore, since 6 pixels in total are involved in the trigger decision, we can estimate roughly that an increase of mean NSB by a factor of $5.4$ (see \autoref{tab:fold_numphot}) due to usage of SiPM instead of PMT will require increase of the trigger thresholds by a factor of 
%$\sqrt{(5.4)^{1/6}}\sim 1.15$. 
%\jsc{this part requires updates. scaling like this was the original idea and it did not work. The current simulations use a  $\sqrt{(5.4)^{1/3}}$ scaling, but that one also have troubles}
%Similarly, in case SiPM camera is equipped with a dichroic window, the smaller increase of NSB by a factor of 1.7 will result in only minor increase of trigger thresholds by  a factor $\sqrt{(1.7)^{1/6}}\sim 1.04$.
The higher level of the NSB fluctuations might require an increase of the trigger thresholds to allow operations with a manageable rate of accidental events. 
However, this effect is also tangled with the difference of the single phe distribution caused by PMT afterpulses and SiPM optical cross-talk as well as with the enlarged pulse integration window. 
Therefore we performed a MC study to derive the expected trigger thresholds for a SiPM camera (see \ref{sec:trig} for details).
Using a MC study, we obtained that a SiPM camera will require a factor of about 1.7 and 1.2 higher thresholds without and with a dichroic window, respectively, to achieve the same level of accidental triggers as for the PMT camera. 
%It should be noted however that those values are obtained assuming a PMT-like single phe distribution and afterpulsing and thus are likely to be overly pessimistic. 
%The precise behaviour will depend on the specific SiPM used and its operational parameters like the overvoltage.

The effect of the increased noise in the images is automatically taken into account, as we simulate the SiPM camera with a higher average rate of NSB phe per pixel. 
Nevertheless, if the noise in the images is highly increased, the image cleaning procedure might need to be revised, in particular the values of the cleaning thresholds need to be increased.
%This is similar to when MAGIC performs observations under Moon presence: the reflected moonlight can be computed as a higher overall NSB level. As an example, when images display 3 (8) times higher NSB due to Moon, a 30 (50)\%  higher image cleaning threshold is required~\citep{Ahnen:2017vsf}.
%To evaluate the effect for observations with a SiPM camera, we simulate a number of "pedestal" events, i.e. events with only electronic noise and NSB photons. 
%We then evaluate the fraction of such "pedestal" events that survive the cleaning.
We optimize the cleaning thresholds based on noise-only events (see \ref{sec:clean} for details). 
Intriguingly, due to the interplay of different effects, the optimal cleaning thresholds for the case of SiPM camera without the filters turn out the same as for PMT camera. 
With addition of the filters the cleaning thresholds can be considerably lowered by a factor of 1.5. 
%despite higher noise level To keep the same fraction of noise islands for the PMT and SiPM case, the standard cleaning threshold  of 6-3.5 phe needs to be increased by 45\% for SiPMa (to 8.8-5.1 phe) and by only 8\% (to 6.5-3.8 phe) in the case of SiPMs equipped with filters.

It is curious that the effect on the trigger thresholds and on the cleaning thresholds during the analysis is different. 
We note however a number of differences between those two stages. 
The MAGIC trigger operates on amplitudes, while the charge extraction used in further  analysis exploits pulse integration. 
The two approaches are not equivalent when time spread of individual phes is present. 
Moreover, the stereoscopic trigger is based on a 2$\times$3NN (next-neighbour) condition with each pixel having a signal above a given threshold.
In contrast, the image cleaning in each telescope is searching for 2NN, 3NN or 4NN combinations with a sum of clipped signals above a given total threshold, applying also different time constraints than a trigger (see \cite{2016APh....72...76A} for details). 
Therefore, the interplay of higher NSB rate with single phe distribution with a smaller tails can have a non trivial, and different  effect on both.

To evaluate the irreducible background by cosmic rays, we simulate the response of the telescopes to proton, helium and electron induced showers. In total, we have simulated 5~M gamma rays, 25~M electrons, 125~M helium 125~M protons between 0$^\circ$ and 35$^\circ$ and 5~M gamma rays, 25~M electrons, 125~M helium  200~M protons between 35$^\circ$ and 50$^\circ$.
We normalise the cosmic-ray proton spectrum to the DAMPE measurement \citep{eaax3793}.
To take into account the effect of helium and higher elements, we apply the same approach as in \cite{2012APh....35..435A}. 
Namely, we normalise the helium to $1.6\times50\%= 80$\% of the proton flux, where the 1.6 factor is the factor derived by \citet{2012APh....35..435A} to correct for elements higher then helium. 
The electron simulations are weighted to the parametrised combination of the \textit{Fermi}-LAT and H.E.S.S. all-electron spectrum applying the model of Eq.~2 from \citet{2021arXiv210505822O}. 
The gamma-ray simulations are normalised to the Crab Nebula spectrum following \cite{2016APh....72...76A}. 

The MC simulations are analysed using the MARS software\footnote{MARS stands for MAGIC Analysis and Reconstruction Software and is a proprietary pipeline of the MAGIC Collaboration.} \citep{2013ICRC...33.2937Z,2016APh....72...76A}. 
%\mdc{What do you want to say, there are anyhow produced within MARS}. 
The MARS software allows to  estimate the nature of the primary particle thanks to a classifier called \textit{hadronness} which is close to 1 for hadrons and 0 for gamma-ray primaries. Further parameters computed by MARS are the estimated energy (derived by using the Random Forest method), as well as the reconstructed event direction in the sky. A proxy for this latter parameter is dubbed $\theta^2$, i.e. the squared angular distance between the nominal and reconstructed source direction.

\section{Performance of MAGIC with a SiPM-based camera}
\label{sec:sensitivity}

\begin{figure*}[t]
% fig updated to 8.3, 5.1 thresholds !
    \centering
    \includegraphics[width=0.49\textwidth]{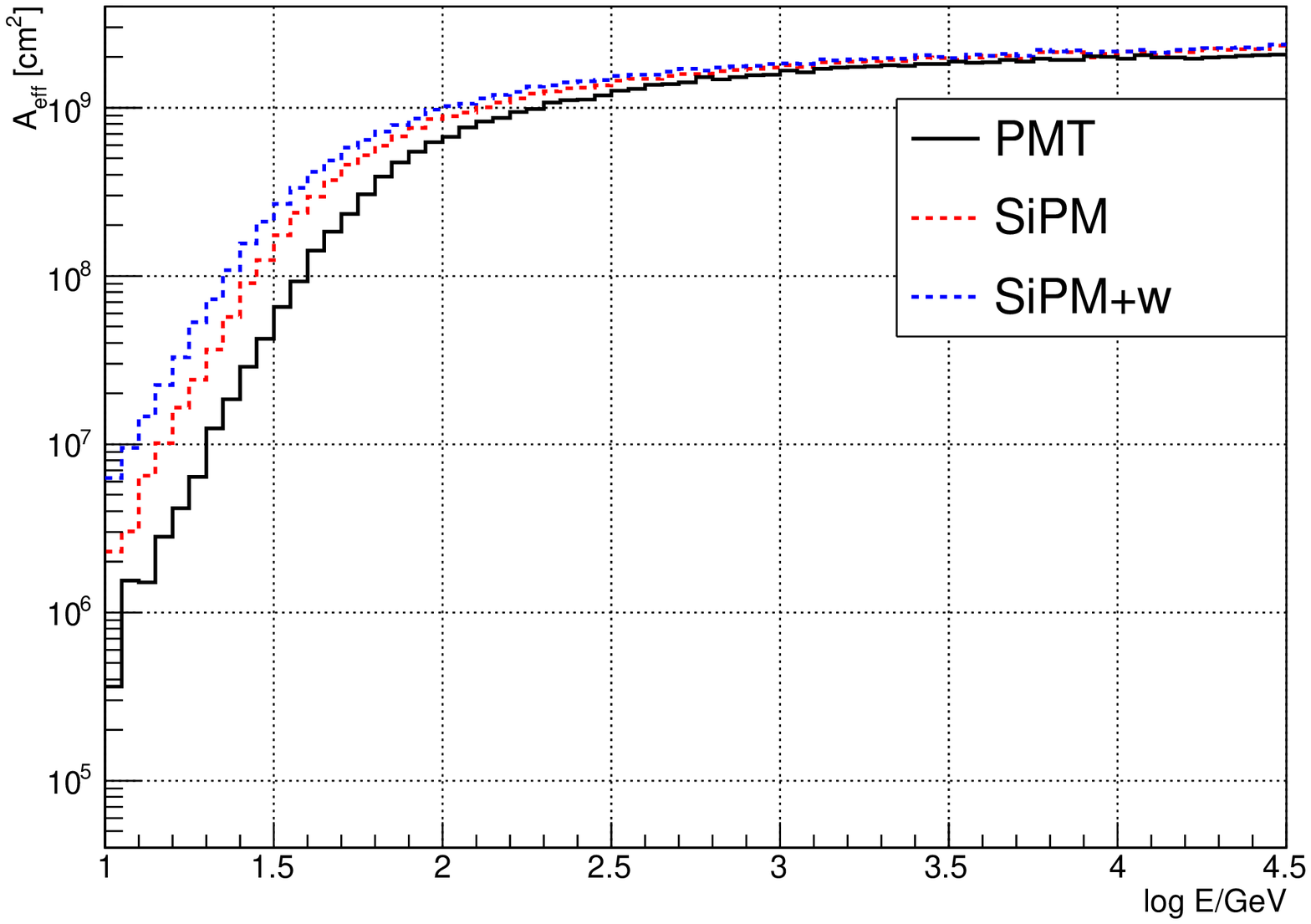}
    \includegraphics[width=0.49\textwidth]{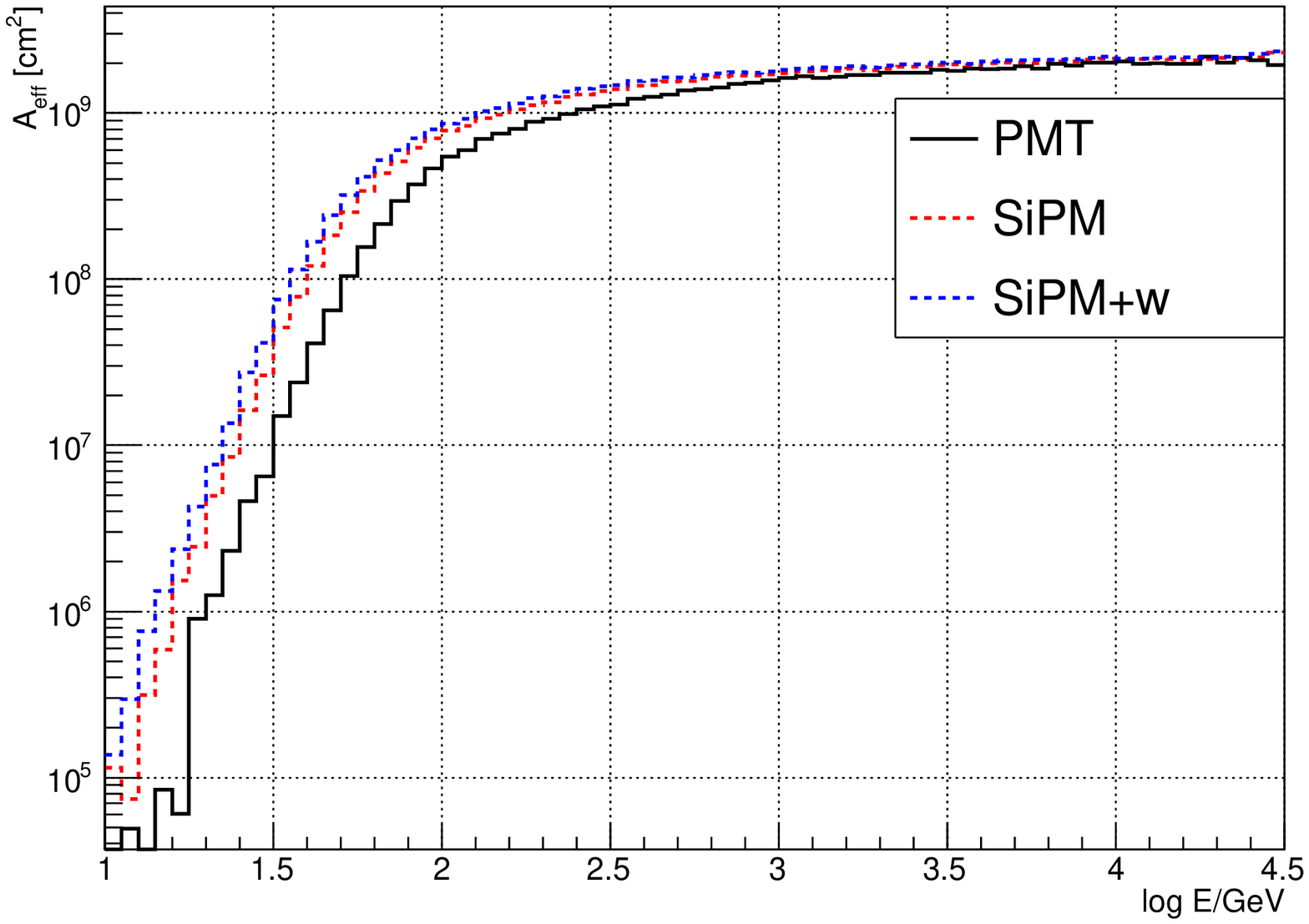} 
    \includegraphics[width=0.49\textwidth]{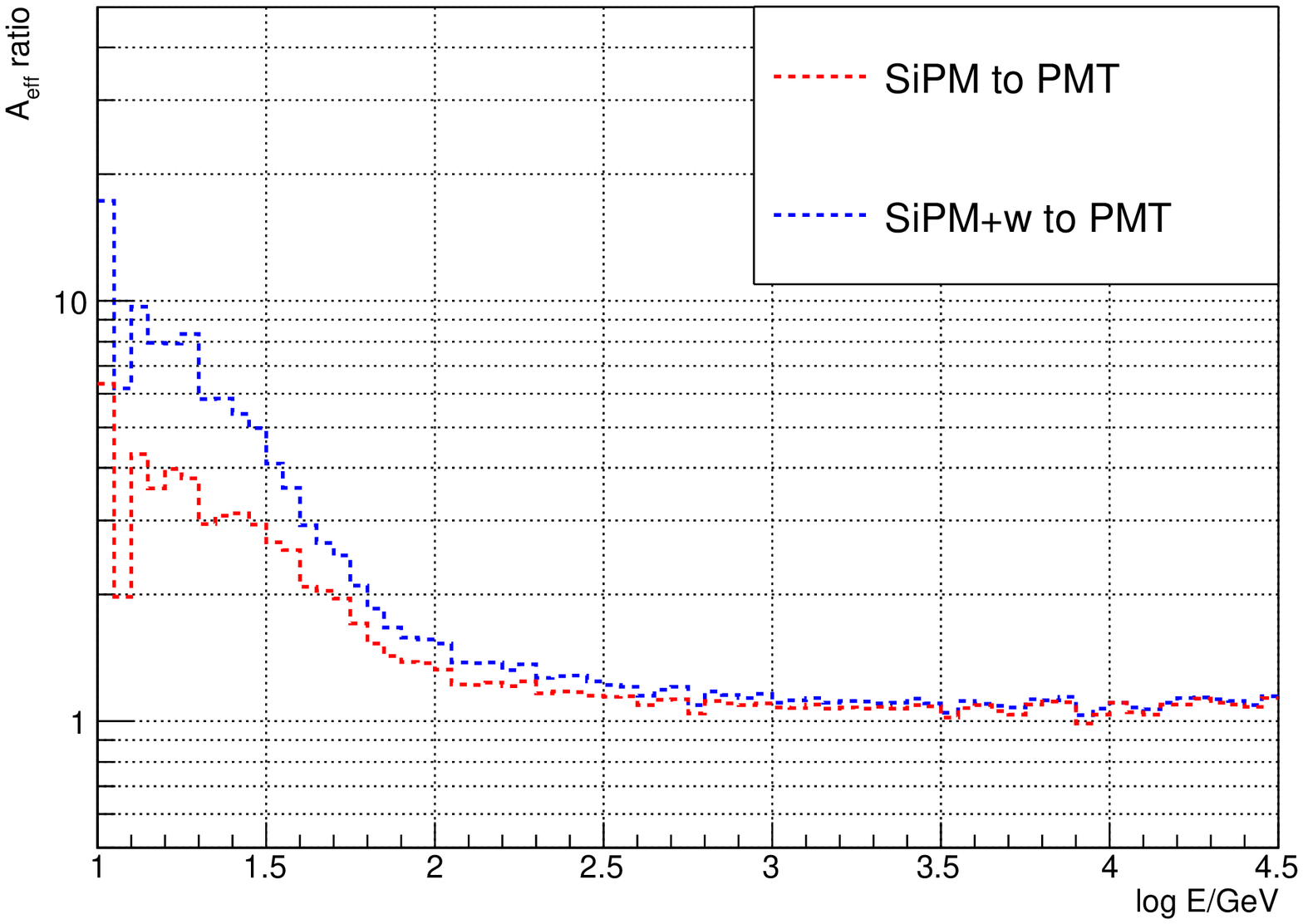}
    \includegraphics[width=0.49\textwidth]{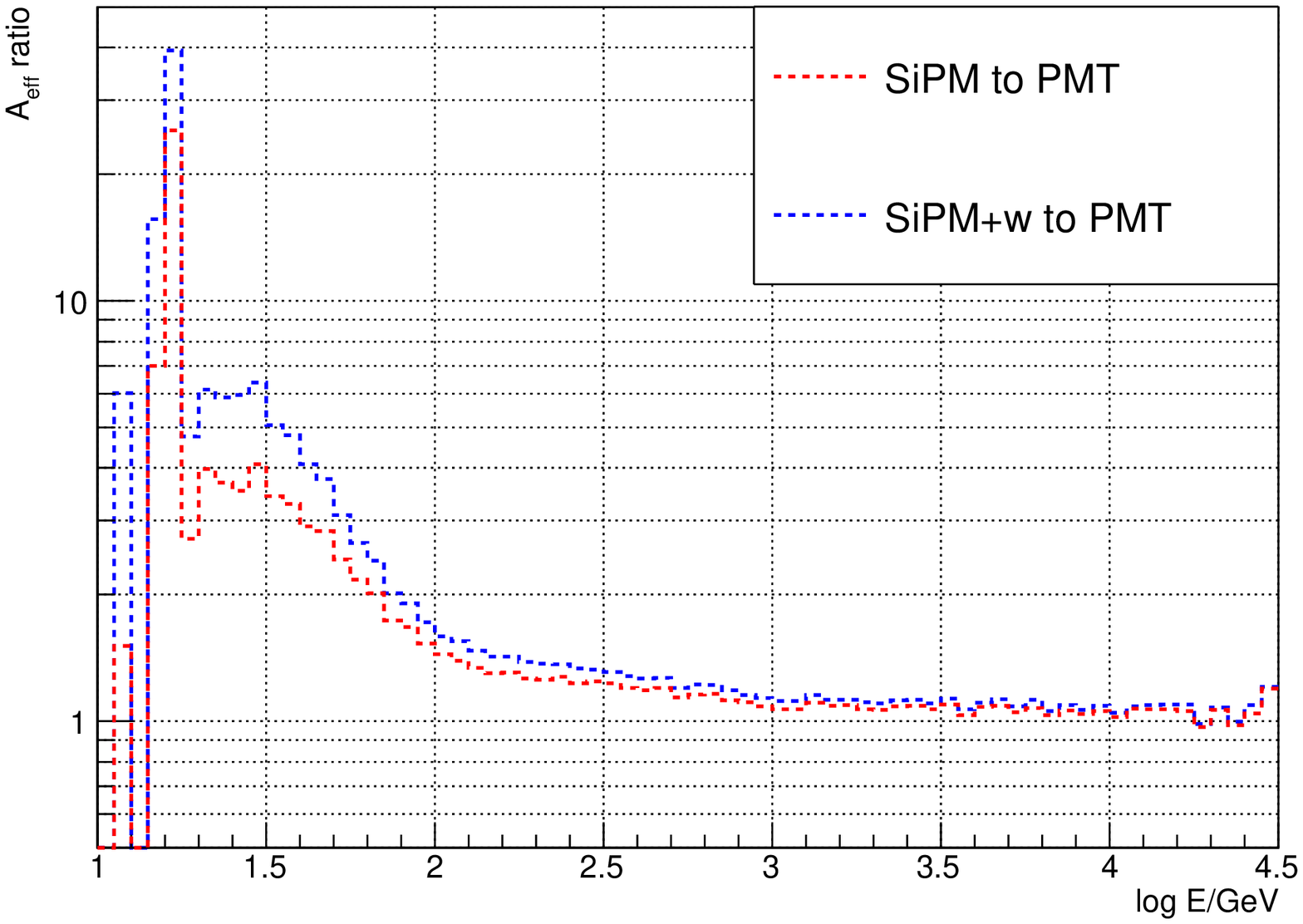}     \caption{
    Averaged effective area for PMT- (black) and SiPM-based camera with (blue) and without (red) filters for low-zenith angle observations. 
    Effective area computed at the trigger level (left panels)
    or at the reconstruction level (right panels). 
    The bottom panels show the ratio of the corresponding effective area to the PMT case.}
    \label{fig:aeff}
\end{figure*}

\paragraph{Effective area with a SiPM-based camera}
To investigate the improvement at the energy threshold of MAGIC,  where the largest effect of the NSB on the event triggering and reconstruction is expected to occur, we have investigated the variation in the collection area. The results are shown in \autoref{fig:aeff}.
At the trigger level (see the top panel),  the higher PDE, despite the increase of the thresholds results in an increase of the effective area at the lowest energies. 
For example at 50\,GeV the collection area is improved by a factor of 2 comparable to that of the PMT camera. 
Moreover, the reduction of noise and thus also the trigger thresholds with the use of filters allows for a further increase of the collection area at the lowest energies accessible with MAGIC. 
At 50~GeV the improvement of filters is a factor $\sim40\%$, resulting in a total improvement of nearly a factor of 3 with respect to the PMT camera. 
We also investigate the collection area at the reconstruction level, namely when both images survive the image cleaning with a minimal size. 
In the case of the PMT camera in the reconstruction, we applied a cut of 50 phe, commonly used in the MAGIC analysis. 
For the SiPM camera, we upscaled this value by 2.2 (without) or 1.7 (with filter),  see \autoref{tab:fold_numphot}.
The improvement of the collection area at the reconstruction stage is comparable to the one at the trigger stage. 
%In this case, the SiPM camera provides a larger collection area than the PMT one, and it is additionally enhanced by the use of a filter. 
%While there is a big improvement below a few hundred GeV visible in the collection area of SiPM cameras to PMT cameras, there is only a very minor gain ($\sim 10\%$ at 30\,GeV) of applying the filters.  

%\medskip
\paragraph{Angular Resolution with a SiPM-based camera}
% figure updated with 8.3 5.1 thresholds 
\begin{figure}[t]
    \centering
    \includegraphics[width=0.49\textwidth]{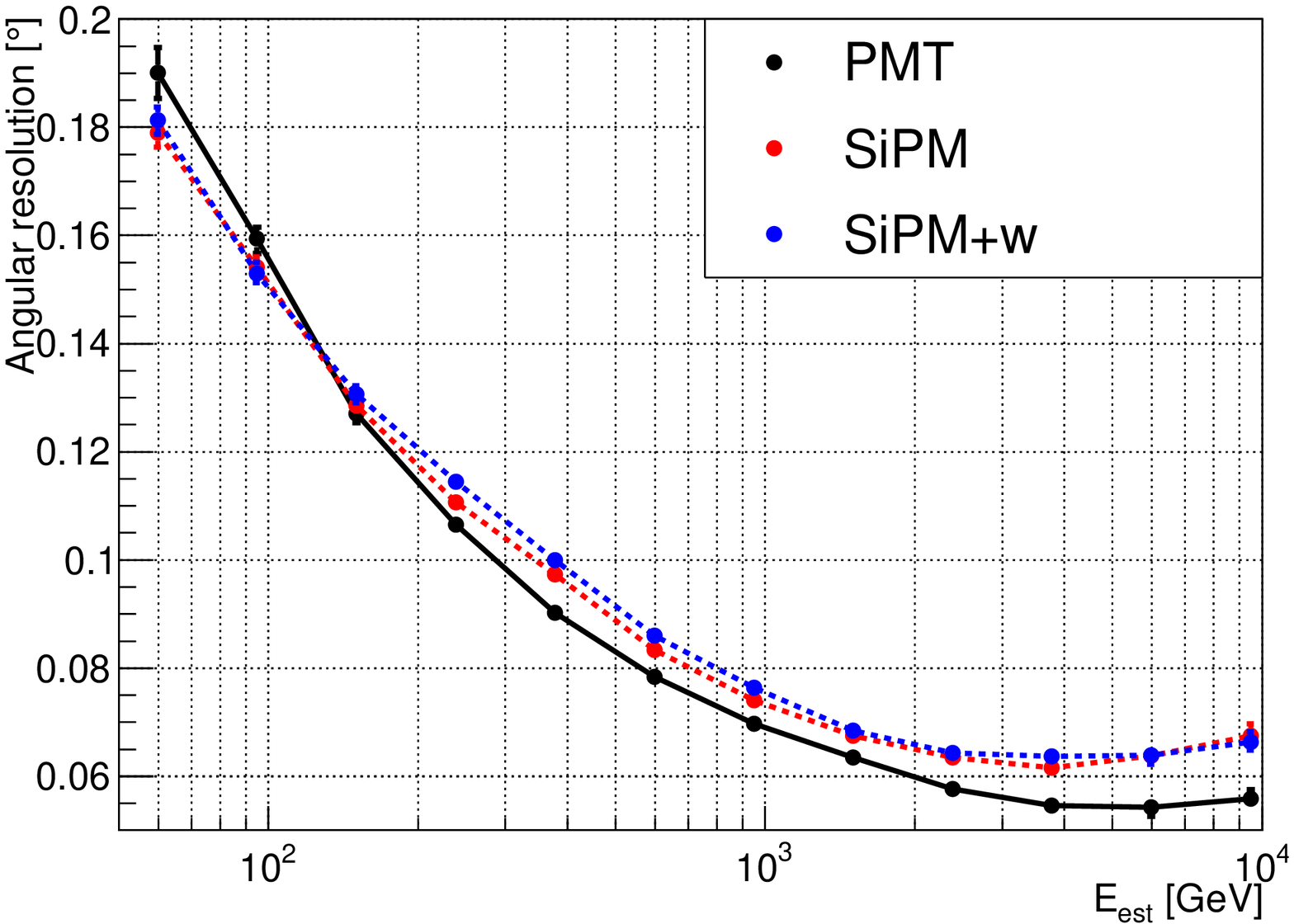}
    \caption{Angular resolution (defined as 68\% containment radius) as a function of the estimated energy for the PMT (black) and the SiPM camera with (blue) and without (red) filters. Simulations at low zenith angles are used. }
    \label{fig:angres}
\end{figure}

A preciser shower reconstruction represents a second possibility by which the improvement of the "cleaningness" of the images may possibly enhance the gamma/hadron separation. 
To study this, we compute the energy-dependent angular resolution for the three simulated camera cases (see \autoref{fig:angres}). 
We apply cuts in minimal image size as done in the effective-area study.  
While the SiPM camera has somewhat worse angular resolution as the PMT one above 200\,GeV, 
below 100\,GeV the angular resolution is slightly improved. There is very little difference in the angular resolution with or without using filters. 
This shows that, except of the lowest energies, the reconstruction of the shower direction is still not strongly limited by the amount of light received by the telescopes, or the "cleaningness" of the image, but by other factors (such as geomagnetic field deflection of the shower, optical PSF of the instrument, pixelisation of the camera, \ldots).  
In fact, the increased collection area of the SiPM camera adds dim images with strong intrinsic fluctuations that are likely less precisely reconstructed, affecting slightly the overall performance.

\paragraph{Sensitivity with a SiPM-based camera}
To evaluate the sensitivity, we analyse the data in bins of estimated energy. 
As for the previous studies of the parameters that govern the instrument performance, we apply a cut in minimal image size with a value scaled to the expected light yield for the SiPM camera. 
In each such bin we optimize the \textit{hadronness} and $\theta^2$ cuts to obtain the best differential sensitivity. 
We scan the values with cut efficiencies from 60\% to 99\% for \textit{hadronness} and from 40\% to 95\% for $\theta^2$. 
%In the case of $\theta^2$ we additionally set the maximum cut value at $0.08^\circ$ - the maximum value allowing 3 symmetric not-overlapping background control regions.
The rate of gamma-ray events is computed directly for a given $\theta^2$ cut. 
However, to improve the statistics for events that contribute to a diffuse background we consider events reconstructed up to $1^\circ$ away from the camera centre, and scale the obtained rate to the much smaller background expected at the solid angle corresponding to the $\theta^2$ value-cut. 
In this scaling we also correct for the drop of the efficiency with the distance from the camera centre by using a fit with a Gaussian function. 
% For the well populated estimated energy bins, the fit is performed in each such bin independently using the least constraining \textit{hadronness} cut. 
The sensitivity is defined as the minimal detectable flux that satisfies three conditions: a $5\sigma$ significance (computed according to Eq.~17 from \citep{1983ApJ...272..317L} assuming three background control regions), an event excess of at least 10 events and above 5\% of the residual background. 
To  use the available statistics of background events in an optimal way, without introducing bias and overtraining the cuts, we divide the individual samples of simulated particles into $N_s=4$ subsamples. 
In the analysis of each subsample, we use $N_s - 1$ remaining subsamples to obtain the optimal cuts, that are then applied to the original subsample. 
All subsamples (processed with the above-described procedure with different cuts) are then stacked together for the final result. 

\begin{figure*}[h!t]
    \centering
    % plots updated to the 8.3 adn 5.1 p.e. thresholds
    \includegraphics[width=0.33\linewidth]{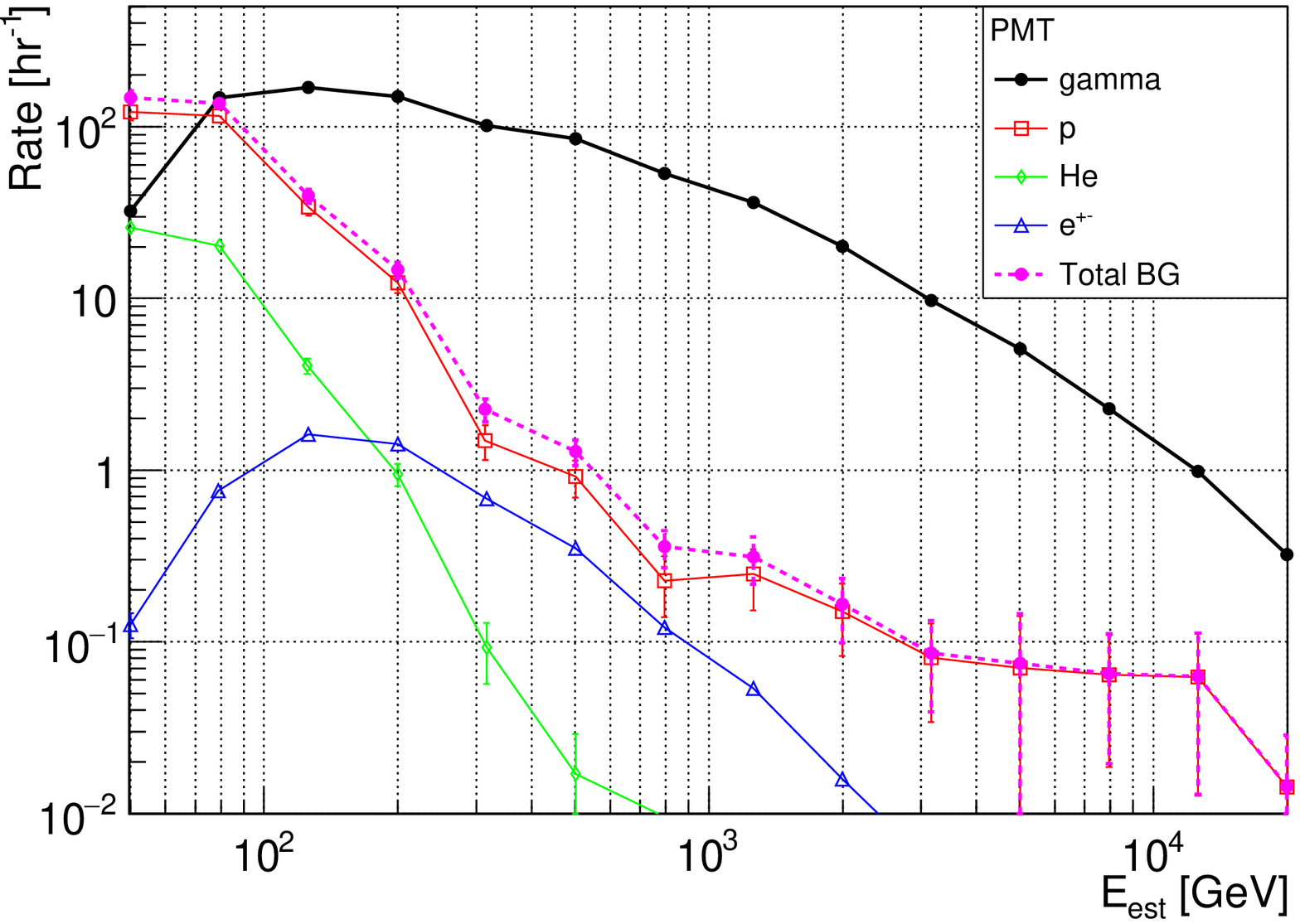}
    \includegraphics[width=0.33\linewidth]{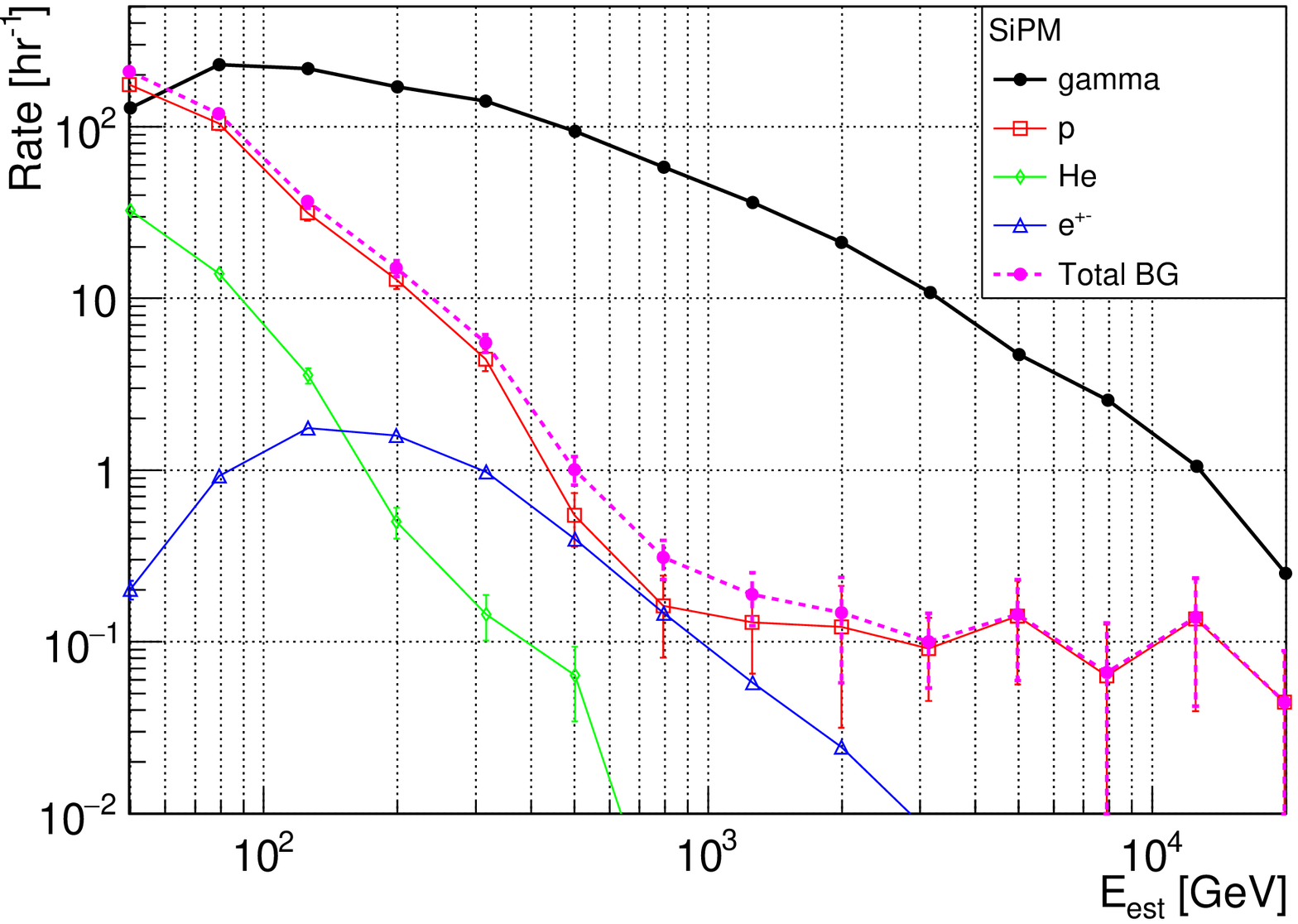}
    \includegraphics[width=0.33\linewidth]{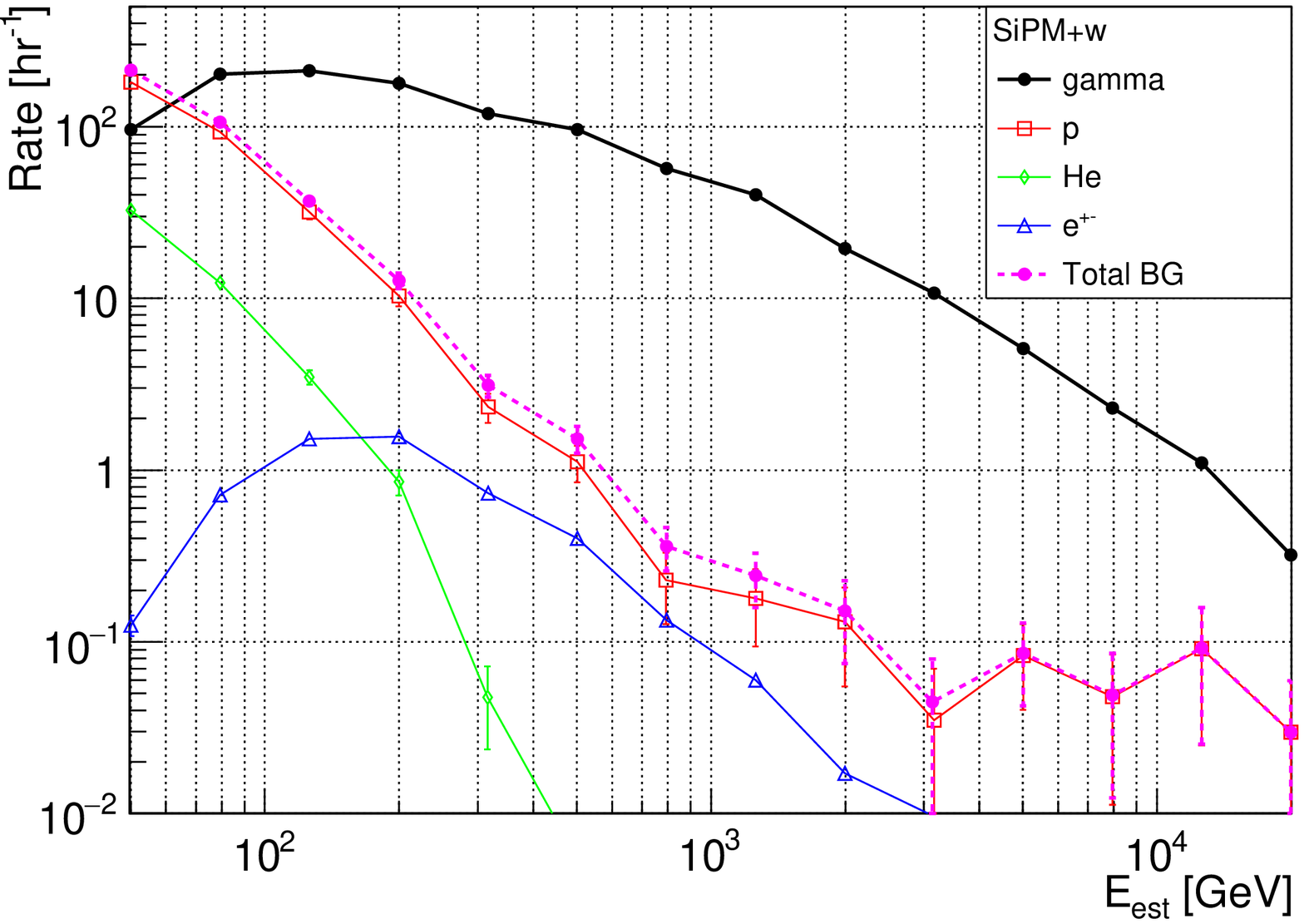}
    \caption{\label{fig:rates}Simulated rates for a SiPM camera for gamma rays (black, solid line, assuming the Crab Nebula spectrum) and the residual background (protons; red squares, helium; green diamonds - including estimation of higher element contribution, all-electrons; blue triangles and total background; dashed line in magenta). 
    The rates are computed with best-sensitivity cuts using MC simulations for low zenith angles. Left panel shows the case of the PMT camera; (middle) case of a SiPM camera without additional filters; (right) case of a SiPM camera with low-pass filter.}
  \end{figure*}

The expected rates for the simulated SiPM camera with and without filters are shown in \autoref{fig:rates}. As expected, the residual background is dominated by protons, with an additional contribution of electrons in the energy range of a few hundred GeV. 
At energies below 100\,GeV the ratio of the gamma-to-background rate is improved when using a SiPM camera. 
The combined effect of the application of filters and re-optimised cleaning thresholds does not change the rates in a qualitative way.

\begin{figure*}[h!t]
    \centering
    \includegraphics[width=0.49\textwidth]{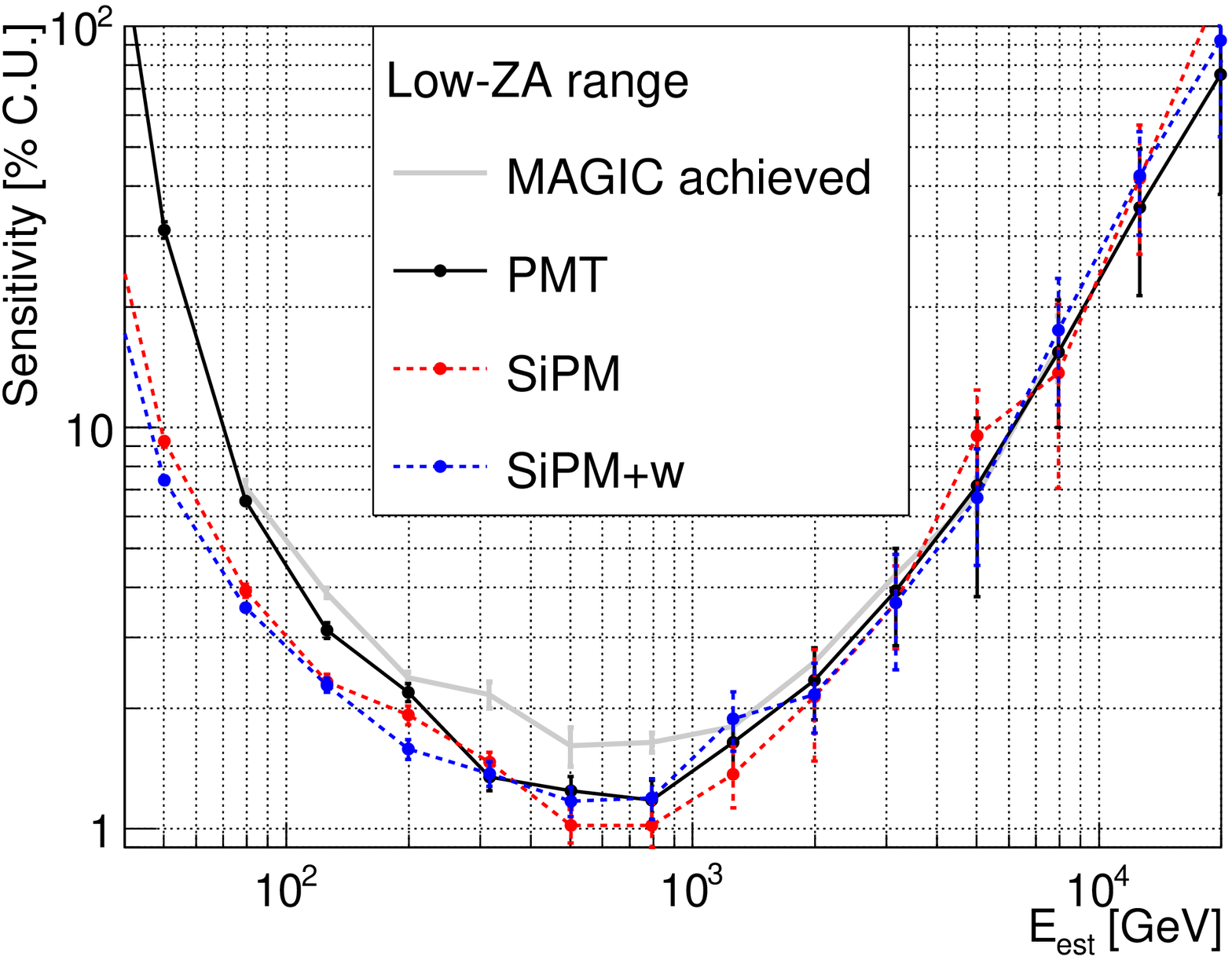}
    \includegraphics[width=0.49\textwidth]{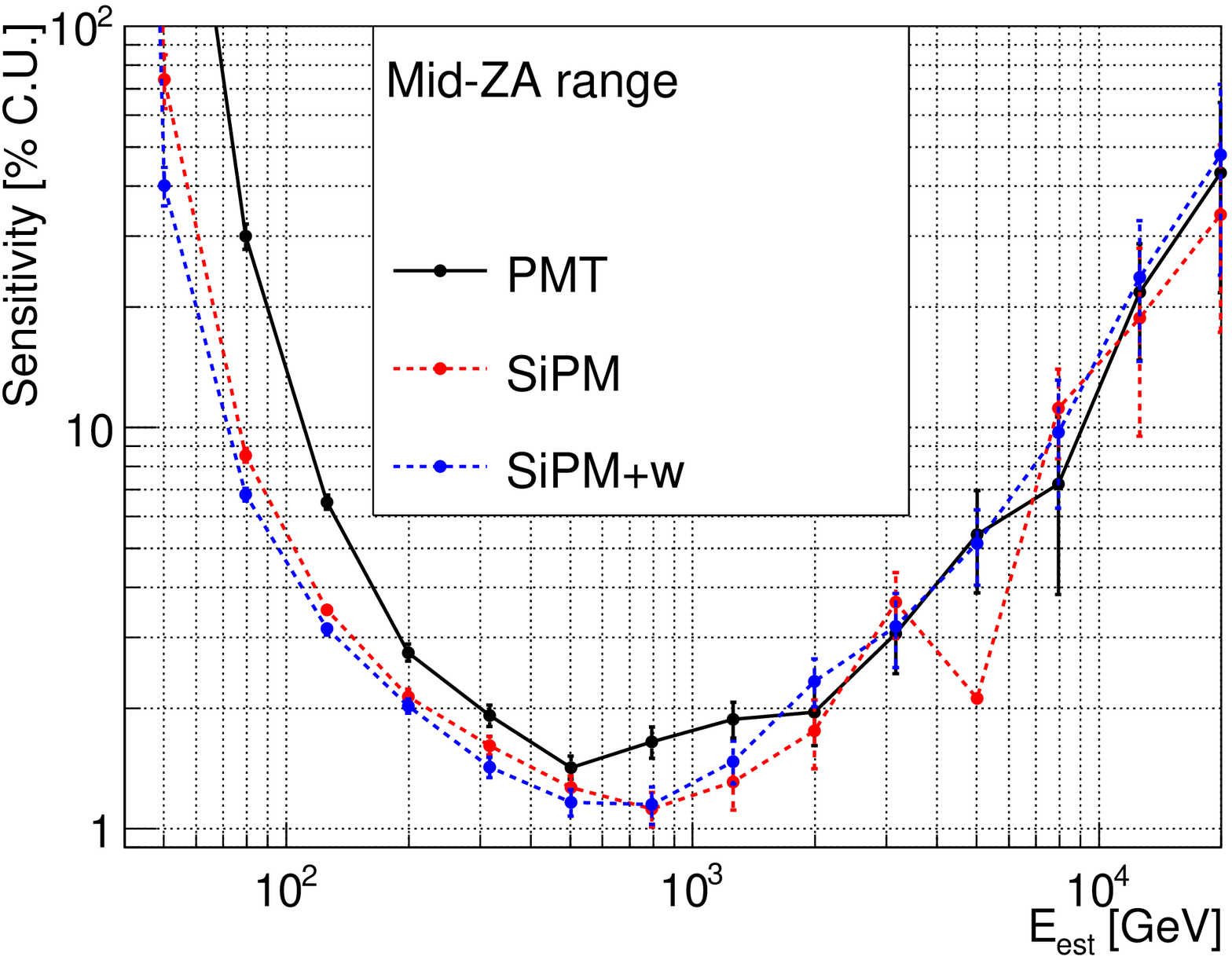}
    \includegraphics[width=0.49\textwidth]{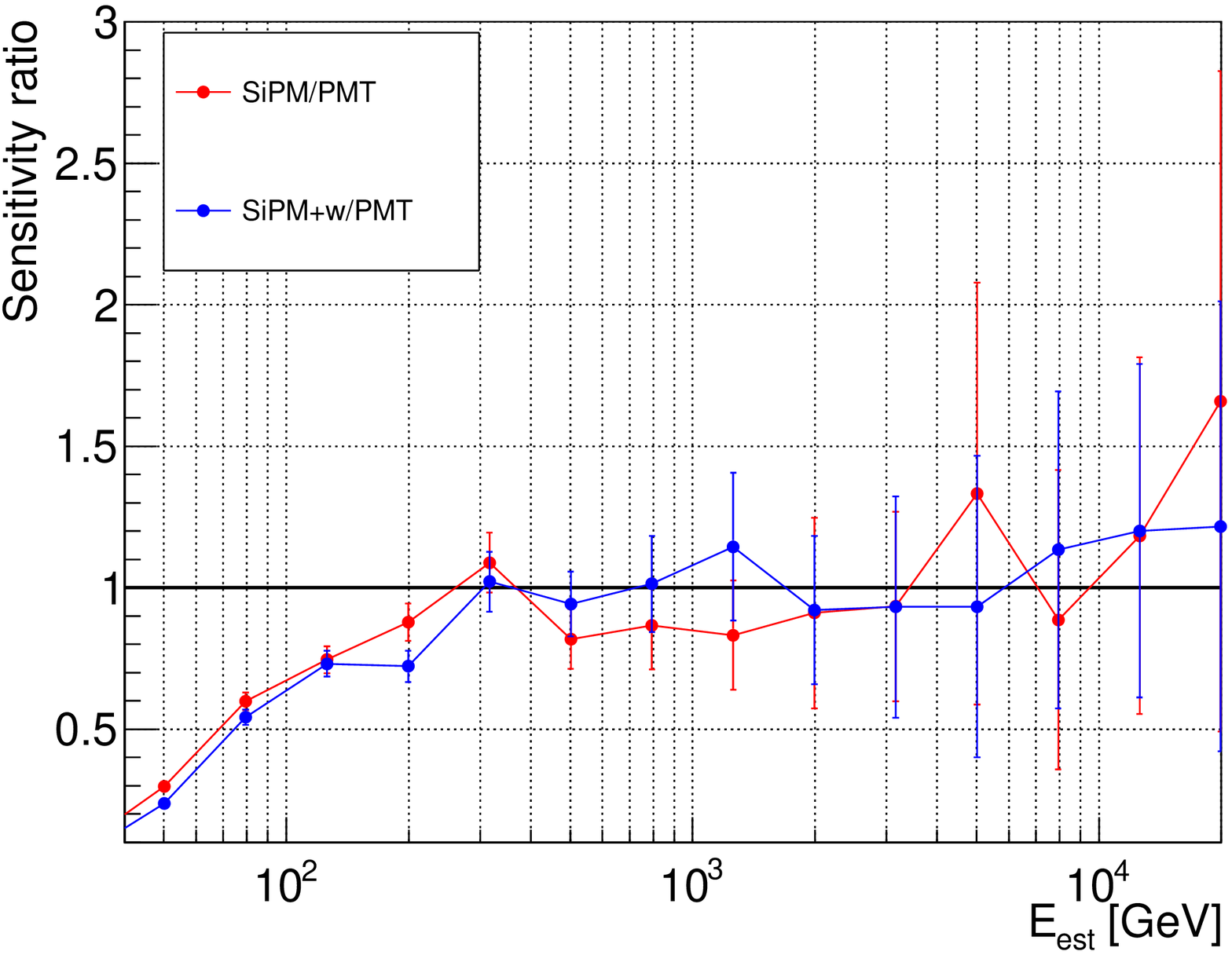}
    \includegraphics[width=0.49\textwidth]{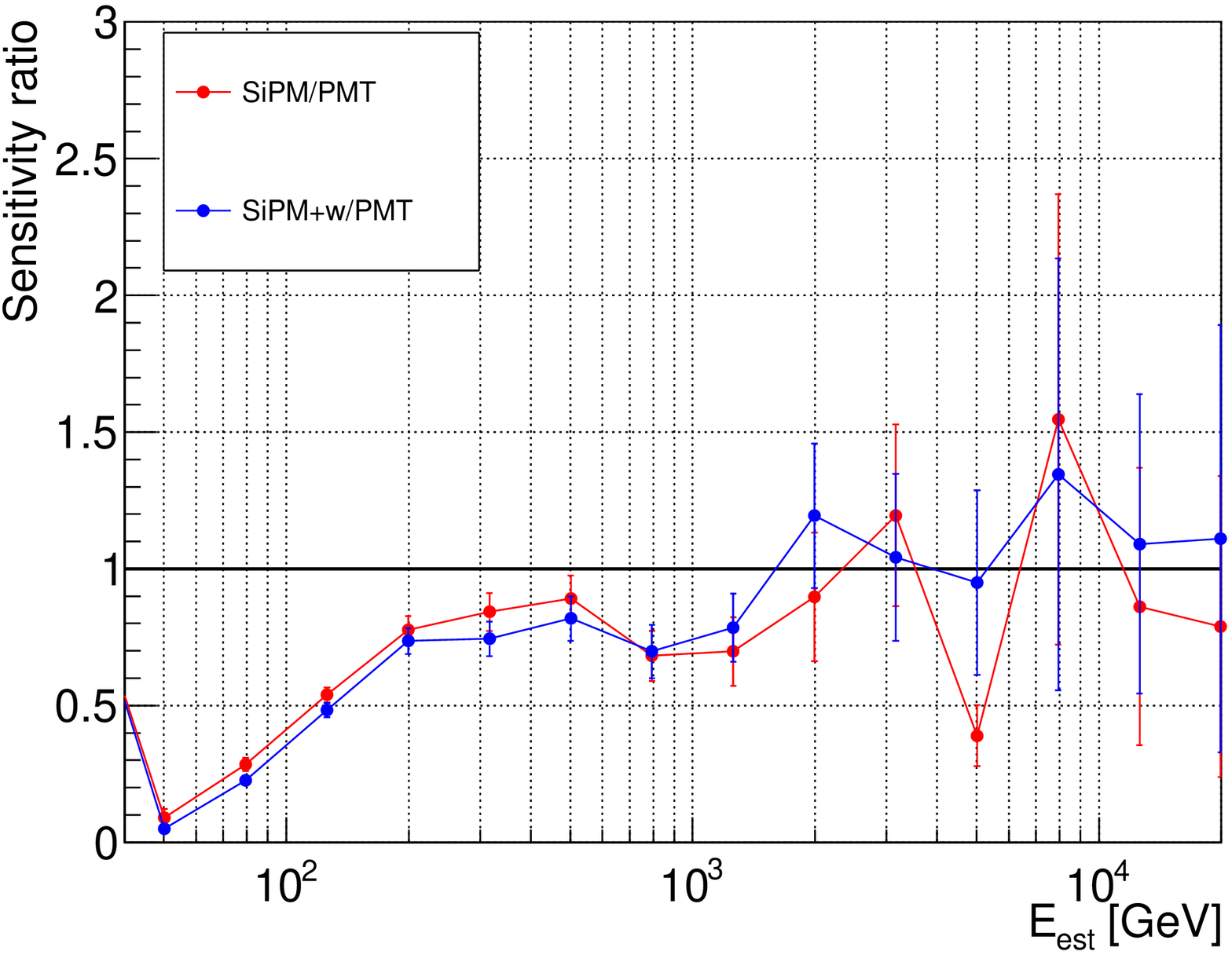}
    \caption{\label{fig:sensitivity}
    Sensitivity (top panels) of a SiPM camera without (red) and with low-pass filter (blue) compared with simulations of a PMT camera (black) and the achieved performance of the MAGIC telescopes (gray, \citep{2016APh....72...76A}). 
    (left) observations at low zenith angles ($5-35^\circ$), (right)  observations at medium zenith angles ($35-50^\circ$). 
    Bottom panels show the corresponding sensitivity ratios with respect to a PMT camera. }
  \end{figure*}
  
%\medskip  
The obtained sensitivities are compared in \autoref{fig:sensitivity}.
The improvement of the SiPM camera is clearly seen in the energy range $\lesssim200$\,GeV for low zenith angles and $\lesssim1000$\,GeV for medium zenith angles. 
%For the low zenith-angle case, the improvement is a factor of two at 50~GeV (i.e., close to the threshold).
For the low zenith-angle case, the improvement is a factor of three at 50~GeV (i.e., close to the threshold).
At the same energy the low-pass filter in front of a SiPM camera gives
an additional small boost of ~20\% in the sensitivity.
However at higher energies the difference in sensitivity by using filters is  smaller, and for the energies $\sim 1$~TeV the obtained sensitivity is actually worse. The result that a low-pass filter has rather small effect on improving the sensitivity is in line 
%While the result might seem curious, it is in line 
with the previous performance study regarding different light illumination levels \cite{Ahnen:2017vsf} showing that moonlight which increases the NSB even by a factor of 5 has negligible ($\lesssim 5-10\%$) impact on the sensitivity in most of the energy range accessible with MAGIC.
In the simulated case of a SiPM camera with and without filters, the application of filters reduces the NSB level by a factor of 3.1 (see \autoref{tab:fold_numphot}), but at the additional price of loosing 20\% loss the light from the air shower. 
%Therefore the expected difference should be even smaller than derived in \cite{Ahnen:2017vsf}. 

Stronger improvement for the medium zenith observations can be caused by the shift of the observed Cherenkov spectrum towards red wavelengths caused by the atmospheric absorption.
The jump in the reported SiPM camera without filters, mid-zenith sensitivity for a 5 TeV point is an artifact caused by low background statistics. 

\section{Discussion}
\label{sec:discussion}

In \autoref{sec:sensitivity} we have seen how the use of SiPMs instead of PMTs may provide a sensitivity boost, which according to our generalized simulations, can amount even to about factor of three at about 50~GeV, regardless the higher sensitivity of SiPMs to the NSB. In this section we discuss the reasons for considering an upgrade of 3$^{\rm{rd}}$ generation IACT cameras to SiPMs, as well as the problems related to the technical implementation.

\medskip
About a factual implementation of a SiPM-based camera, we start by recalling that our results are not based on a full-scale design and simulation of the camera, 
%\ca{but we did simulations so is this correct?} \md{we did not simulate geometry, light funnel, amplification stage, summation stage, daq, trigger}
and thus this is just a preliminary orientating work. To accurately address the performance change with a fully SiPM-based camera, one should in detail define the actual SiPM to be used, the SiPM summation electronics, at least. 
In the generalized SiPM simulations that we applied in this work we limited to taking care of those effects at the level expected from the available devices. 
%All in all, the specific aspects discussed above, concurring to the precise determination of the sensitivity, should not provide a performance worse than what we predict, wherefore 
Therefore we may claim our results conservatively as follows: at low energies SiPMs would provide an improved performance.

Regarding the addition of a modified plexiglass window, with dichroic properties, we also mention that, although it looks like the improvement of the performance of such a window is not very high.
The biggest improvement is expected in terms of collection area, at the lowest energies, however due to simultaneous increase of background rate it does not affect sensitivity much (note that at the lowest energies the sensitivity is limited by the signal-to-background systematic condition, which means that the sensitivity scales with the background to excess   rather than square root of background to excess). 
Such a gain in the collection area can be considered a benefit in case of some scientific targets, e.g.: observations of pulsars (where signal-to-background condition does not apply), or very short transients. 
Nevertheless, it seems that the additional effort, cost and complexity of installing wavelength filters is not justified performance-wise. 
On the other hand such filters also do not significantly degrade the performance, thus can be considered a viable solution if the particular hardware implementation could be problematic e.g. due to pile-up of copious NSB photons. 
Therefore, we have contacted two companies about the technical feasibility of a dichroic window that cuts wavelengths above 550~nm. It is a substantial technological effort but we understood it is affordable. While obtaining an optimal cut ($>95$\% transmission below 550~nm and $<5$\% transmission above 550~nm) is obtained through the deposition of a series of dichroic layers, more complex is the substrate choice. The technology of shaping the light transmission through transparent materials is very advanced nowadays and achieved through repeated deposition of coating layers whose global interference pattern allows to create a dichroicity figure customizable to the percent level\footnote{For a fixed incident direction.}. The layer deposition process is normally achievable on small surfaces at custom level or large surfaces at industrial level. Possibly arrangements in tiles can be considered, such as that used in MAGIC or VERITAS for special observation modes~\cite{Guberman:2015rsa}. Glass layers, that would be the best substrate, would be too heavy and fragile, considering that the camera is subject to repentine acceleration.  Plexiglass or similar materials are lighter and more flexible, however they suffer ageing from UV light, which is used in the dielectric layers deposition. Also, the deposition of homogeneous layers over a wide area is technologically a challenge, albeit possible for some companies. More easily obtained is a camera composed of few large tiles of dichroic pads held together by a thin frame. 
%All in all, we were proposed two solutions for an affordable budget of about 20~k\texteuro.

\medskip
Although we cannot support this claim without a full simulation, we are of the opinion that our results obtained for MAGIC are roughly valid also for the other two $3^{rd}$ generation IACTs currently operating: H.E.S.S. and VERITAS.  
The boost in sensitivity at low energies would generally improve the prospects of detection for targets in which the low-energy signal is more relevant, however, it might not be sufficient to argue for a full camera upgrade for H.E.S.S. and VERITAS before the advent of CTA. However, we argue that for MAGIC it could be a viable opportunity. MAGIC was designed and constructed to be the lowest energy-threshold instrument of its kind \citep{Doro:2012nd}. 
The reason is that MAGIC has a vantage outlook to the Northern extragalactic sky, where a broadband multiwavelength coverage is facilitated. 
%The reason is that MAGIC has a vintage point of view to the  Extragalactic sky, being located in the Northern Hemisphere. 
The gamma-ray signal from distant extragalactic sources suffers strong absorption in the Extragalactic Background Light fields with increasing energy if the gamma-ray, hence the need for lowest possible energy thresholds. There is a second reason why an upgrade of the camera of MAGIC could be considered. The MAGIC telescopes are located very close to where the fist telescope of the Cherenkov Telescope Array North (CTA-N) is built, the LST1 telescope~\cite{2023arXiv230612960P}. %\citep{Cortina:2019juz}. 
Additional three LST telescopes are currently under construction. LSTs are 23-m class telescopes with similar parabolic shape as the MAGIC telescopes, but with twice larger surface area that 
allows sampling the fainter signals of lower-energy events. 
Currently, LST1 is operating both as a standalone instrument and in joint observation with MAGIC, mostly because the commissioning of IACTs is facilitated in stereoscopic mode, because a shower location is better triangulated, but also because the joint observation allows for an improvement of about 50\% in the LST1 performance~\cite{2019ICRC...36..659D}.
An improvement of the MAGIC sensitivity at the lowest energies would go into the direction of matching MAGIC's performance towards LST-1. We remark however that there are no plans for the long term of MAGIC telescopes being part of the CTA-N array, so this solution could prove only temporary.

%\medskip

\section{Summary and Conclusions}
\label{sec:conclusion}
In this work we have discussed the possible upgrade of a current-generation IACT camera from PMTs to SiPMs (\autoref{sec:sipm}). 
%We have found that:
\begin{enumerate}
    \item We have found that the spectral Cherenkov photons distribution in \autoref{tab:ch_peak} and \autoref{fig:ch_spectrum} at ground depends little on the energy and more on the zenith angle (except for the total yield) and this is a measurable effect when folded with the instrument spectral yield (see  \autoref{sec:yield4});
    \item There are several factors that affect the IACT acceptance to different frequencies of the Cherenkov signal, as well as to the  background and the NSB (see \autoref{sec:yield3}). Most of these have approximately a 'gray' (wavelength independent) response (see \autoref{fig:window}), such as the mirror reflectivity, the plexiglass window protecting the camera and the light funnels on top of the photosensors. The stronger wavelength-dependent response comes from the PDE of the photosensors. We have shown exemplary PDEs of PMTs and SiPMs showing that the typical sensitivity of SiPMs to red photons is higher w.r.t. that of PMTs (see \autoref{fig:pde})
    \item We have found that by replacing PMTs with SiPMs in the camera and comparing the light yield, we have found that SiPMs would allow for a factor 2.2 to 2.5 higher signal (\autoref{tab:fold_numphot}) regardless the energy of the primary. This strong signal boost is also accompanied by an even stronger factor of 5.3 of more light from the NSB (\autoref{fig:compare_numphot}) that requires higher suppression of the NSB both at the trigger and analysis level (optimised trigger and cleaning thresholds).
    \item We have investigated the performance gain through end-to-end Monte Carlo simulations and found that a SiPM-based camera would grant a factor of about three higher sensitivity at 50~GeV (for low zenith-angle observations), with decreasing improvement for higher energies (\autoref{fig:sensitivity}). This is due to the fact that the higher-energy showers are optimally reconstructed even with a less sensitive photosensor. At medium zenith angles some improvement is seen up to 1\,TeV energies.  
    \item We have investigated the possibility of reducing the sensitivity to the red part of the spectrum, specially intense in the NSB. We have found that by replacing the plexiglass with a dichroic window, we could still gain a factor $\sim1.8$ more signal from the showers and only a factor of  1.7 more light from the NSB. However, this solution only moderately improves the sensitivity with respect to a neutral plexiglass in combination with a SiPM camera (\autoref{fig:sensitivity}). This is because the sensitivity of MAGIC is only affected at very high NSB level.
    \item The simulations show that a SiPM camera will also display slightly  better at low energies, angular resolution (\autoref{fig:angres}), however at higher energies the effect is inverted. 
    \item Additionally (especially if equipped with filters) the SiPM camera will provide a better effective area at low energies (\autoref{fig:aeff}). 
    \item We have discussed the technical implementation of such cameras in \autoref{sec:discussion}, finding that our estimates could be quite accurate or, in worst case, conservative.
\end{enumerate}

Replacing the PMTs of the MAGIC cameras with SiPMs would be a costly intervention, but it could be a valid investment for MAGIC in prolonging its scientific competitiveness and in connection to CTA.

\paragraph{Acknowledgments.} \begin{small}
The authors warmly thank the anonymous referees who carefully checked the first version of the manuscript and suggested further checks will lead to improvement of our results. The authors would also like to thank the MAGIC Collaboration for allowing to use their simulation and analysis tools for performing this study. We especially acknowledge A. Hanh for useful comments on the manuscript. and D. De Paoli, R. Rando and M. Mariotti for useful discussions. MD acknowledges funding from Italian Ministry of Education, University and Research (MIUR) through the "Dipartimenti di eccellenza” project Science of the Universe. JS is supported by Narodowe Centrum Nauki grant number 2019/34/E/ST9/00224.
\end{small}

\bibliographystyle{unsrtnat}
\bibliography{pde}

  \appendix
  \section{Tunning of threshold parameters}\label{sec:trig}
  The higher light yield of the SiPM devices, as well as the differences in the low-level performance parameters (e.g. pulse width, single phe distribution / afterpulsing) are expected to affect the level of the noise induced by NSB. 
  This in turn will determine the operation parameters (threshold), which will have an impact on the achieved low-energy performance.  
  In this appendix we explain the tuning procedure of the simulated trigger threshold parameters and cleaning thresholds in the subsequent analysis
  \subsection{Trigger threshold}
\begin{figure}[t]
      \centering
      \includegraphics[width=0.49\textwidth]{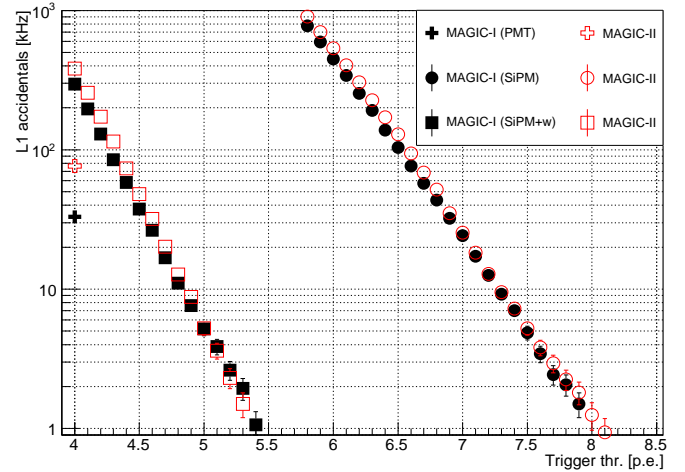}
      \caption{The expected rate of accidental triggers for SiPM camera (circles) and SiPM with filters camera (squares) as a function of trigger threshold (amplitude in phe) for MAGIC-I (black filled markers) and MAGIC-II (red empty markers). 
      For comparison simulations of PMT-camera at the operational threshold of MAGIC are shown with crosses. 
      }
      \label{fig:trig_acc}
  \end{figure}

  In order to determine the optimal trigger thresholds we simulated 100 000 events without Cherenkov photons. 
  For each of them we simulated the response of the telescopes to the NSB within a 160 ns window, and computed number of events producing a single telescope trigger in such a time window. 
  We convert the fraction of triggers $f_t$ into accidental rate $r_a$ as $r_a = f_t/$(160\,ns) and plot it in Fig.~\ref{fig:trig_acc}.
A similar accidental rate of a SiPM camera without filters requires increase of the thresholds by $\sim70\%$, while with the addition of filters by only $\sim12\%$.

  \subsection{Cleaning threshold} \label{sec:clean}
\begin{figure}[t]
    \centering
    \includegraphics[width=0.49\textwidth]{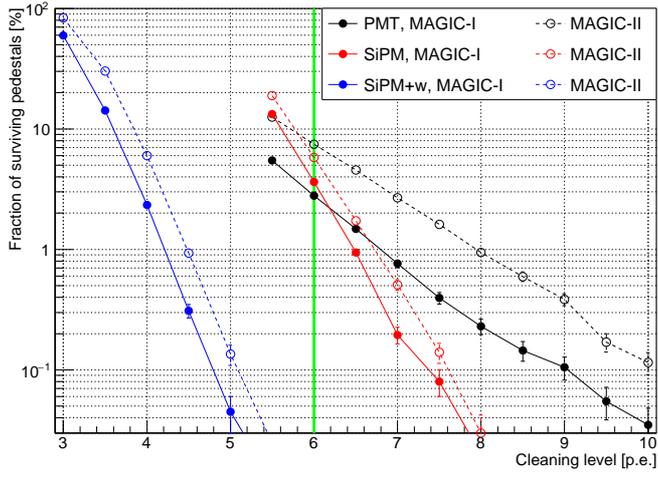}
    \caption{Fraction of noise-only events surving image cleaning with a particular cleaning level threshold. 
    SiPM without (red) and with (blue) the wavelength filters is compared with the PMT camera (black). 
    MAGIC-I is shown with filled circles and solid lines, while MAGIC-II with empty circles and dashed lines.
    Thick vertical green line represent the standard cleaning threshold of PMT camera. }
    \label{fig:surv_ped}
\end{figure}

This study is using the standard MARS cleaning, the so-called Sum cleaning with different time and charge thresholds for 2NN (next-neighbour), 3NN and 4NN combinations \citep{2016APh....72...76A}. 
The individual NN charge thresholds are normalized to a common cleaning level expressed in phe.
With a different level of noise the optimal cleaning level for the analysis might be also affected. 
To test this we generated and calibrated 20 000 events without Cherenkov photons, with only NSB and electronic noise.
Next we applied different cleaning levels, for each of them checking the fraction of events surviving the cleaning. 
Such a fraction can be interpreted as the maximum fraction of shower events affected by an artificial islands in their image solely due to NSB fluctuations, and thus should be kept low. 
In the case of PMT simulations the cleaning normalization level of 6 phe results in $\sim 5\%$ of empty events surviving the cleaning. 
Fig.~\ref{fig:surv_ped} reports the results for SiPM camera without and with the filters.
The optimal cleaning level for the SiPM camera is similar to the one of PMT camera.
With the addition of the wavelength filters instead a factor of 1.5 lower cleaning threshold can be achieved. 
\end{document}